\documentclass[12pt, a4paper, twosides]{article}
\usepackage[T1]{fontenc}
\usepackage[utf8]{inputenc}
\usepackage[english]{babel}
\usepackage{graphicx}
\usepackage{mathtools}
\usepackage{enumitem}
\usepackage[top=1 in,bottom=1in, left=1 in, right=1 in]{geometry}
\graphicspath{{./}{./figures/}}
\usepackage{authblk} 
\usepackage{verbatim}
\usepackage[page, title]{appendix}

\usepackage{textcomp,gensymb}
\usepackage{booktabs,caption}
\usepackage{subfigure}
\usepackage{graphicx}
\usepackage[colorlinks]{hyperref} 
\usepackage[dvipsnames]{xcolor}
\usepackage{booktabs}
\usepackage{fancyhdr}
\begin{document}

	\title{Elastic disk with isoperimetric Cosserat coating}
	
	\author[1]{ Matteo Gaibotti }
	\author[1]{ Davide Bigoni \thanks{Corresponding author: Davide Bigoni 
			fax: +39 0461 282599; phone: +39 0461 282507; web-site:
			http://www.ing.unitn.it/$\sim$bigoni/; e-mail:
			bigoni@ing.unitn.it.}}
	\author[2]{ Sofia G. Mogilevskaya }
	
	\affil[1]{DICAM, University of Trento, via Mesiano 77, I-38050 Trento, Italy}
	\affil[2]{Departement of Civil, Environemental and Geo- Engineering, University of Minnesota, 500 Pillsbury Drive S.E. Minneapolis, MN 55455-0116, USA.}
	\affil[  ]{\text{e-mail: matteo.gaibotti@unitn.it, bigoni@ing.unitn.it, mogil003@umn.edu}}

	\date{}
	\maketitle
	\thispagestyle{fancy}
	
	\vspace{5 mm}
	\begin{center}
	{\it 
	Dedicated to \\ 
	Professors Natasha and Sasha Movchan on the occasion of their 60th birthday}
	\end{center}
		\vspace{5 mm}

	\begin{abstract} 
		
		A circular elastic disk is coated with an elastic beam, absorbing shear and normal forces without deformation and linearly reacting to a bending moment with a change in curvature. The inexstensibility of the elastic beam introduces an isoperimetric constraint, so that the length of the initial circumference of the disk is constrained to remain fixed during the loading of the disk/coating system. The mechanical model for this system is formulated, solved for general loading, and particularized to the case of two equal and opposite traction distributions, each applied on a small boundary segment (thus modelling indentation of a coated fiber). The stress fields, obtained via complex potentials, are shown to evidence a nice correspondence with photoelastic experiments, {\it ad hoc} designed and performed. The presented results are useful for    the design of coated fibers at the micro and nano scales.

	\end{abstract}

	\section{Introduction}	
	
Several technologies involve coating of the surface of a bulk material with a thin layer made up of another material.  This technique was developed in order to enhance electrical conductivity \cite{li2011electrically,liang2007atmospheric,qi2018versatile}, achieve electrical insulation \cite{norizuki2021fabrication,shen2019silica},  protect from heating \cite{liu2017heat,zakirov2018study}, enhance biocompatibility \cite{oshida1994fractal,soldatova2020composite,suzuki2009ultraviolet},  increase strength \cite{tokariev2013strength} or wear, fatigue, or corrosion resistance \cite{jinlong2020wear}. 

A coating layer diffuses the load on an attached solid in a nonlocal way,  introduces a characteristic length, and deeply affects the mechanics of the coated object. As a consequence, a strong research effort has been devoted to the modelling of coatings and to the analysis of associated problems. 
In the framework of the nonlinear theory of elasticity, the coating may be modelled as a surface possessing a membrane stiffness \cite{gurtin1975continuum}, possibly enhanced with a flexural and torsional stiffness \cite{steigmann1999elastic,steigmann1997plane}. Applications of these theories have been presented to bifurcation \cite{bigoni2018bifurcation,dryburgh1999bifurcation} and wave propagation \cite{gei2002Vibration} in coated elastic blocks. 

Within the realm of linear elasticity, the coating has been idealized as an elastic shell (becoming a curved beam in a two-dimensional formulation) \cite{benveniste1989stress,benveniste2001imperfect}. In  composites,  the presence of a film separating two elastic media has been reduced to a thin interface, suitably describing the contact conditions. This interface model has been used to estimate the effective mechanical properties of composites with inclusions  \cite{benveniste1985effective, benveniste1989stress,bigoni2002statics} and for applications in thermal conductivity \cite{miloh1999effective}.

The advent of nanotechnologies has strongly fueled the  development of interface models. These have been used to estimate thermo-mechanical properties of nano-composites, of mono-layered-graphene based materials \cite{suk2010mechanical,wei2019nanomechanics}, to analyze the interaction between nano- inhomogeneities \cite{mogilevskaya2008multiple} and the mechanics of reinforcements, such as coated fibers \cite{han2018local,mogilevskaya2018elastic} or nanoplatelets embedded in a core matrix \cite{baranova2020analysis, mogilevskaya2021use}. 
The presence of coatings or thin layers strongly influences failure mechanisms, such as crack nucleation and propagation inside the coating/substrate \cite{hutchinson1991mixed,jorgensen1995cracking,suo1989steady},  
or delamination induced by 
mechanical/thermal mismatch and residual stresses existing between coating  and substrate \cite{beuth1992cracking,gioia1997delamination,he1994crack,hu1988decohesion,jensen1990decohesion,jensen2002numerical,suo1990interface,yu2001edge,yu2003delamination}, or by curvature changes or buckling \cite{lu2011surface,wu2013buckling,zhuo2015mode}.
	\begin{figure}[hbt!]
		\centering
		\includegraphics[keepaspectratio,scale=1.10]{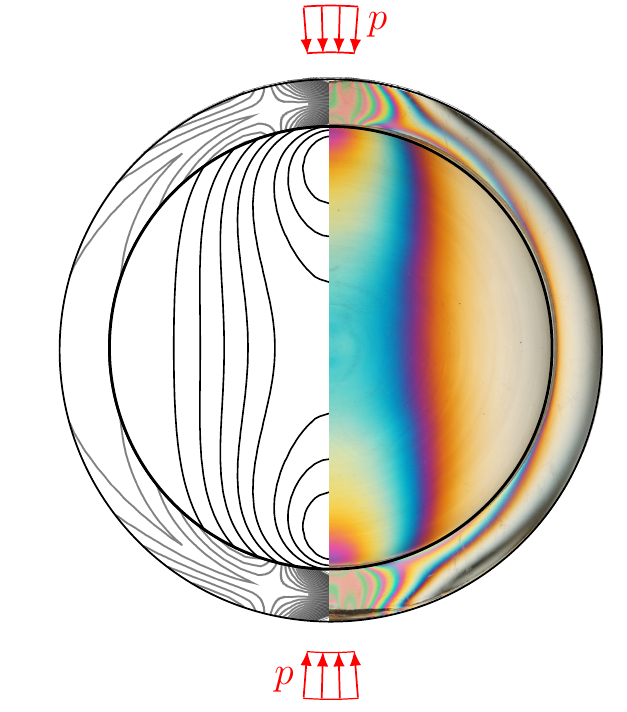} 
		\hspace{0.5 pt}
		\caption{
		Qualitative comparison between the in-plane deviatoric stress   ($\left|\sigma_{I}-\sigma_{II}\right|/\mu$, analytically evaluated with the disk/coating model introduced in the present article) and the photoelastic fringes resulting during a diametral compression test of a coated disk. The sample before testing is reported in the upper part of Fig. \ref{sample_Geom}. The applied external load distribution $p$ is modelled with a Fourier series expansion, enhanced with the Lanczos smoothing method (Section \ref{lanczosSec}).
		}
		\label{coatedDisk}
	\end{figure}

In the present article a circular disk made up of a (linear, isotropic) elastic material is analyzed, 
covered with a \lq beam-like' model of coating, of the type introduced by Benveniste and Miloh \cite{benveniste2001imperfect}. 
The beam model is assumed to satisfy inextensibility, so that an isoperimetric constraint is introduced of the Cosserat type, as the beam transmits bending\footnote{
More precisely, an Euler-Bernoulli beam is an example of constrained Cosserat material, whereas a Timoshenko beam corresponds to the unconstrained Cosserat case, called also micropolar unconstrained theory.}. 
The cylindrical geometry considered here may be important in view of the development of nanowire technology (where a coated cylinder is obtained, when a nanowire is grown inside a single-wall carbon nanotube \cite{wang2021surface}). 
We show that, when the exterior of the coated disk is loaded by a generic (but self-equilibrated) force distribution, the 
problem can be analytically solved via complex potentials, thus obtaining the displacement, strain, and stress fields within the disk, together with the axial and shear forces and bending moment in the coating.

The analytical result is complemented with the explicit treatment of the case in which the coating/disk system 
is subject to a load distribution modelling two equal and opposite concentrated forces. This load corresponds to that applied during a  nanoindentation test (often  
performed on nanofibers to measure their mechanical properties \cite{qi2003determination}) and may easily be reproduced experimentally. On this vein, we have designed two coated 
disks, manufactured (with a CNC engraving machine) from a single block of polymethyl methacrylate, so that the bonding between the coating and disk is perfect and residual stresses are absent.
The samples have been tested 
in a circular polariscope and the results strongly support the coated disk 
model,  
so that the photoelastic fringes are very well captured by the elastic solution, as anticipated in Fig. \ref{coatedDisk}. The figure 
shows the strong effect of the coating, where the stress distribution 
is typical of that forming inside a curved beam subject to bending moment. 

The article is organized as follows. After the coating/disk system is modelled in Section \ref{coatdiskCmplx}, the 
equations governing its behaviour are presented (Section \ref{coatdiskSol}). The special case of loading consisting of two equal and opposite radial force distributions applied to a small area is solved in Section \ref{PhSec} and the solution is compared with photoelastic experimental results in Section \ref{PhEvSec}.

	\section{Modelling of the complex coating/disk \label{coatdiskCmplx}}
	
	A (linear and isotropic) elastic circular disk is examined, enclosed in a perfectly-bonded annular elastic beam, which obeys the Euler-Bernoulli model, so that it is axially inextensible, unshearable and reacts linearly to a curvature variation through the development of a bending moment. The mechanical model of the coating/disk system follows from the combination of the separate equations holding for its two components, which are introduced below.

	\subsection{The coating, a circular annular beam}
	
\begin{figure} [hbt!]
	\centering
		\includegraphics[keepaspectratio]{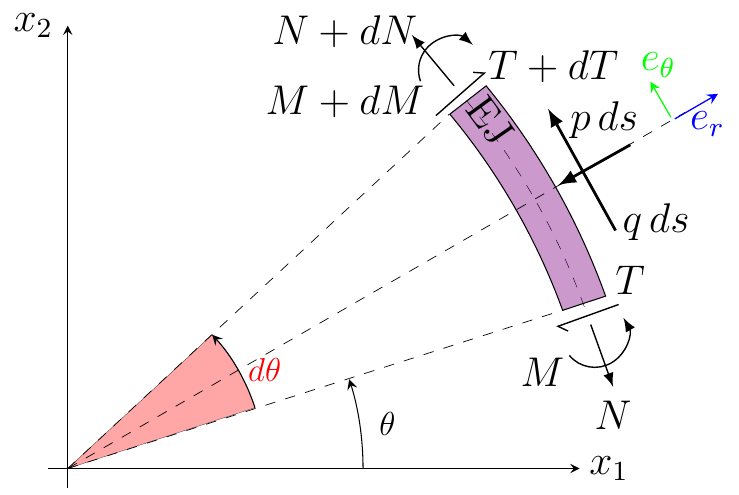}
		\caption{The coating of the disk is modelled with an inextensible and unshearable elastic beam (with bending stiffness $EJ$). The beam introduces an isoperimetric closed contour $L$, so that the perimeter of the coating cannot change its total volume, when a load of components $p$ (radial) and $q$ (tangential) is applied. This load generates internal normal and shear forces, $N$ and $T$, and bending moment $M$. }
		\label{circularbeam}
	\end{figure}

	A circular Euler-Bernoulli (unshearable) elastic beam of bending stiffness $EJ$ and radius $R$ is 
	considered. The center of the circle is located at the origin of the Cartesian coordinate system with
 axes $x_1$ and $x_2$ and the polar system $r$ and $\theta$ (the latter assumed positive when counterclockwise), equipped with the two radial and circumferential unit vectors ${\bf e}_r$ and ${\bf e}_\theta$ (Fig. \ref{circularbeam}). The beam is subjected to radial and tangential forces, respectively $p$ and $q$, that will be applied by both the external environment and the elastic disk. The internal forces along the beam, comprised of normal and shear components, $N$ and $T$, as well as bending moment $M$, satisfy {\it equilibrium}
	\begin{equation}
	\label{equilibrium}
	\frac{dN}{ds}+\frac{T}{R}=-q,  ~~~
	-\frac{dT}{ds}+\frac{N}{R}=-p, ~~~
	\frac{dM}{ds}=-T, 
	\end{equation}
	where $ds=Rd\theta$ is the elementary arclength defined anticlockwise accordingly with $\theta$. 
	
\begin{figure}[hbt!]
		\includegraphics[scale=0.97]{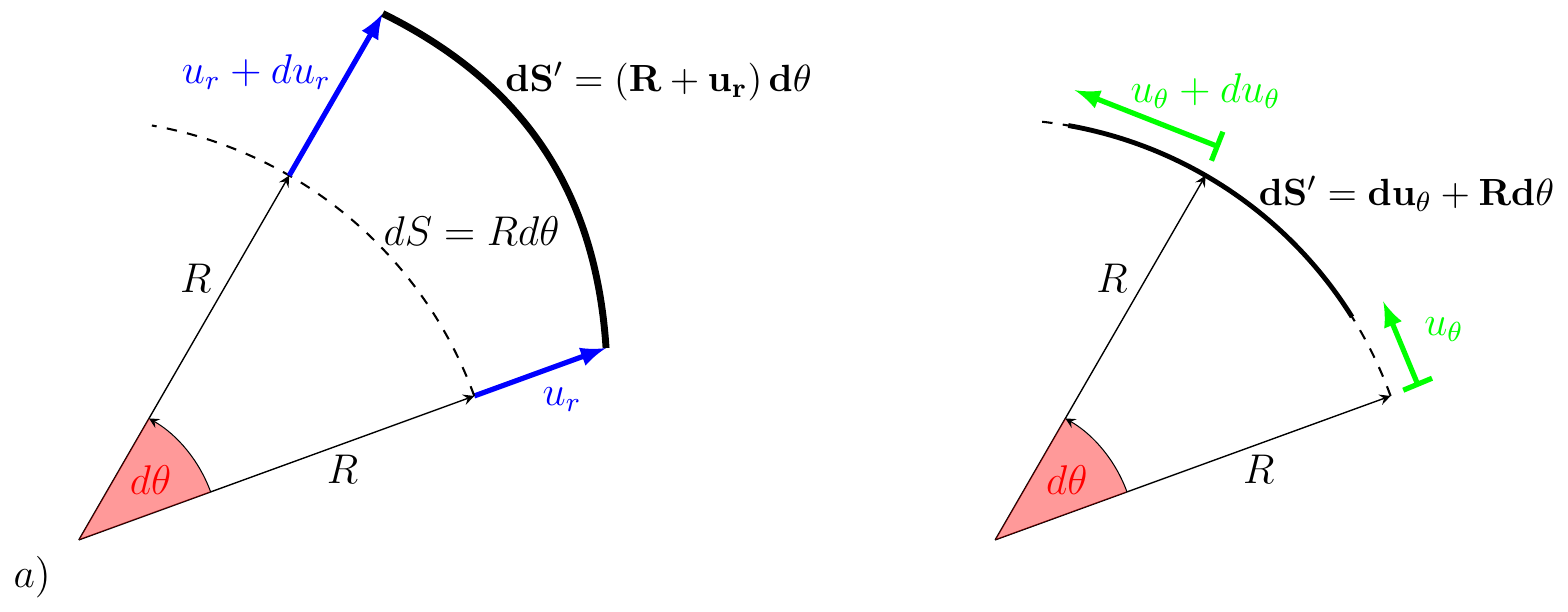}\\
		\includegraphics[keepaspectratio]{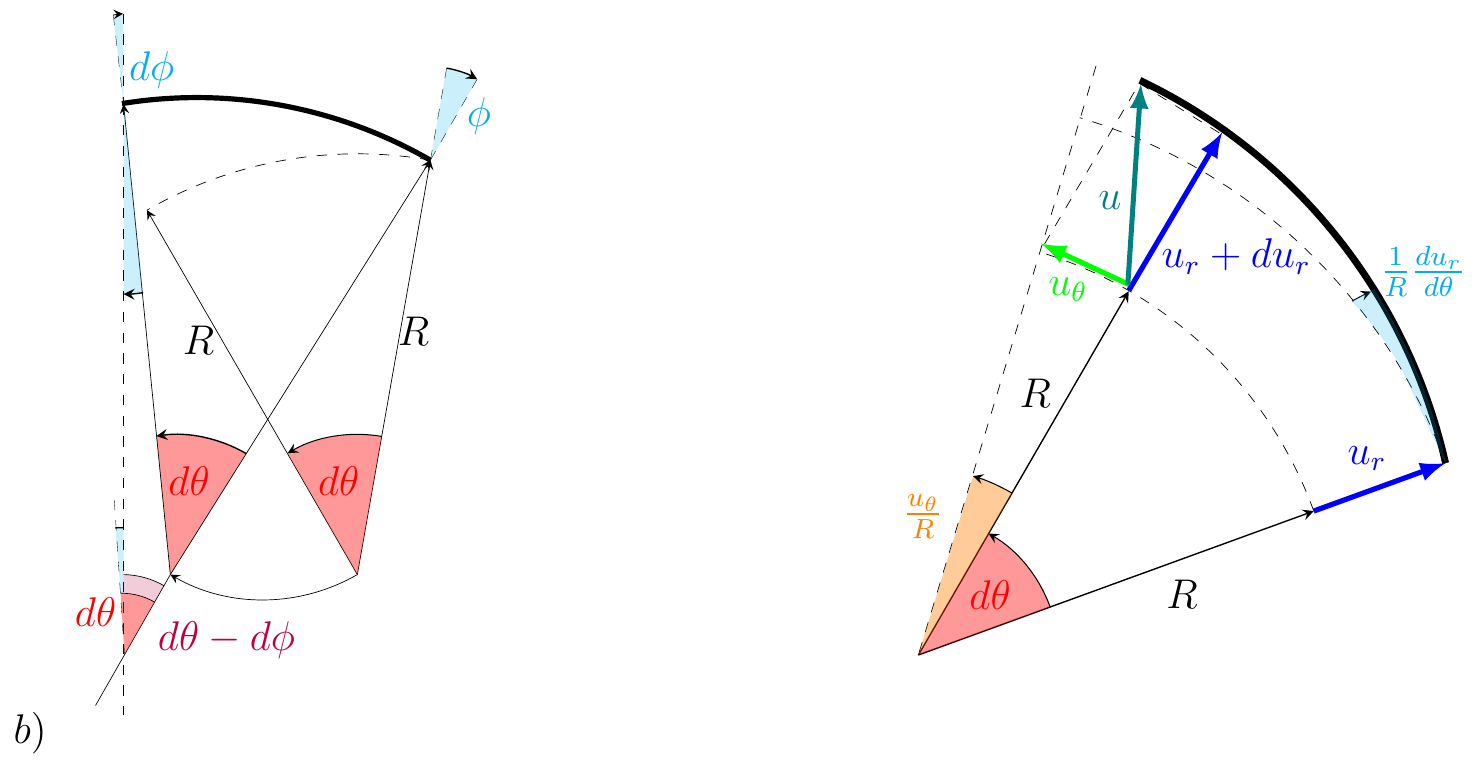}
		\caption{ Kinematics of a circular beam modelling the coating of the elastic disk. a) On an infinitesimal beam element the sum of both radial $u_r$ and tangential $u_\theta$ displacement generate a total length variation $\epsilon$ described by Eq. \ref{cinema}. b) The  rotation $\phi$ of a cross section of the beam is related to $d\theta/d\phi$, the tangential displacement $u_{\theta}$, and the increment $du_{r}$.}
		\label{circularbeam2}
	\end{figure}

	The {\it kinematics} of the beam is defined by a radial $u_r$ and tangential $u_\theta$ displacement and a rotation $\Phi$, the latter positive when clockwise. Pure kinematic considerations, sketched in Fig. \ref{circularbeam2}, show that the axial deformation $\epsilon$, the rotation $\Phi$, and the curvature $\chi$ (the latter additional and opposite to the natural curvature $1/R$ of the circle) are given by \cite{timoshenko1950strength} 
	\begin{equation}
	\label{cinema}
	\epsilon = \frac{u_r}{R} + \frac{d u_\theta}{ds},  ~~~  \Phi = -\frac{u_\theta}{R} + \frac{d u_r}{d s},  ~~~ \chi = - \frac{d \Phi}{d s}.
	\end{equation}
	
	The beam represents an example of constrained Cosserat solid, capable of reacting to a change in curvature with a bending moment according to the linear {\it constitutive equation}  
	\begin{equation}
	\label{costi}
	\chi = - \frac{M}{EJ}.
	\end{equation}

	It is assumed now that this beam be 
	{\it axially inextensible}, so that $\epsilon = 0$, namely, 
	\begin{equation}
	\label{inextensibility}
	\frac{du_{\theta}}{d\theta}=-u_{r} , 
	\end{equation}
	a relation which introduces a local isoperimetric constraint for the area enclosed inside the annular region, which perimeter cannot change. 
	Here,  differently from the mathematical \lq Queen Dido problem', the curve enclosing the area is an elastic beam, equipped with a finite bending stiffness and subjected to external loads.
	
	Equations \eqref{equilibrium}--\eqref{costi}, determining the mechanics of an annular beam under quasi-static loading, can now be particularized for the inextensibility constraint $\epsilon=0$ to hold, so that they become functions of the radial displacement $u_r$ only, as follows. 
	\begin{itemize}
		
		\item Kinematics governing the displacements and rotation, $u_r$, $u_\theta$, and $\Phi$:
		\begin{equation}
		\label{cinemm}
		u_\theta = - \int u_{r}(\theta) \, d\theta, 
		~~~
		R\Phi = -u_\theta + \frac{d u_r}{d \theta}, 
		~~~
		\chi = - \frac{1}{R} \frac{d\Phi}{d\theta}.
		\end{equation}

		\item Constitutive equations for the internal forces $M$, $T$, and $N$:  
		
		\begin{equation}
		\label{azione}
		\begin{array}{ll}
		\displaystyle \frac{M}{EJ} = \frac{d^2u_r}{ds^2}+\frac{u_r}{R^2}, 
		\\ [5 mm]
		\displaystyle
		\frac{T}{EJ} = - \frac{d^3u_r}{ds^3}-\frac{1}{R^2}\frac{du_r}{ds}, 
		\\ [5 mm]
		\displaystyle
		\frac{N}{EJ} = -R \left( \frac{d^4u_r}{ds^4}+\frac{1}{R^2}\frac{d^2u_r}{ds^2} +\frac{p}{EJ} \right).
		\end{array}
		\end{equation}
		
		\item Differential equation for the radial displacement $u_r$: 
		\begin{equation}
		\label{differenzialona}
		\frac{d^5 u_r}{d\theta^5}+
		2 \frac{d^3 u_r}{d\theta^3} +
		\frac{d u_r}{d\theta} = 
		\frac{R^4}{EJ} \left(q - \frac{dp}{d\theta}
		\right).
		\end{equation}
	\end{itemize}
	
	Equations \eqref{inextensibility} and \eqref{differenzialona} define the so-called $\textit{inextensional shell type}$ interface model introduced by 
	Benveniste and Miloh  \cite{benveniste2001imperfect}, their equations (2.15)$_3$ and (2.15)$_4$ with $N=3$, in the particular case when the radius of curvature is constant. More in detail, equation (4.9) in \cite{benveniste2001imperfect} coincides with the above equation \eqref{differenzialona}.
	
	\subsection{The coated disk}
	
\begin{figure} [hbt!]
	\centering
		\includegraphics[keepaspectratio]{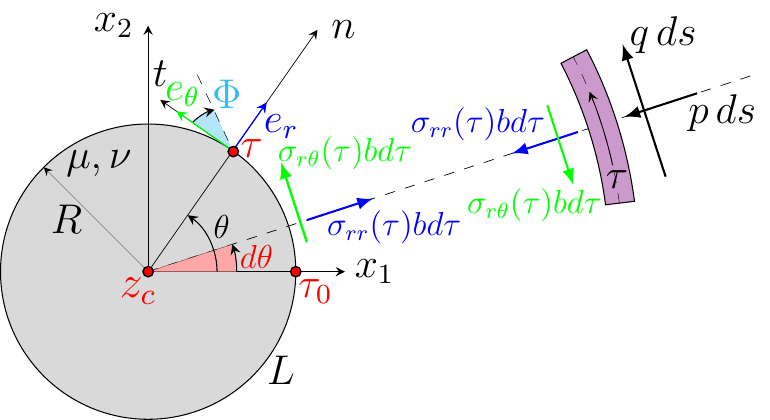} 
		\caption{Elastic disk coated with an inextensible and unshearable elastic beam (with bending stiffness $EJ$).
		} 
		\label{coat22}
	\end{figure}

	A two-dimensional (thus subject to conditions of either plane stress or plane strain) elastic disk of radius $R$, made up of an isotropic material (defined by a shear modulus $\mu$ and Poisson's ratio $\nu$), 
	is coated with the above introduced beam  (Euler-Bernoulli, axially inextensible and unshearable, Fig. \ref{coat22}).  
	The coating is perfectly connected to the disk and hence conditions of continuity of displacement at the contact between the two impose the following conditions 
	\begin{equation}
	\label{continuity}
	u^{\left(b\right)}_{r}=u^{\left(d\right)}_{r}\big{|}_{r=R}, ~~~
	u^{\left(b\right)}_{\theta}=u^{\left(d\right)}_{\theta}\big{|}_{r=R} \, ,
	\end{equation}
	where $(b)$ and $(d)$ denote the \lq beam' and the \lq disk', respectively. These  superscripts will be in the following avoided, as the distinction between the two will be not needed, as they have been assumed to coincide.

	The axial inextensibility of the coating enforces the isoperimetric constraint (\ref{cinemm})$_1$ on the points of the boundary $L$ of the disk
		\begin{equation}
		\label{inextensibilityInt}
		u_{\theta} =-\frac{1}{R}\int_L u_{r} \, ds,
		\end{equation}
defining a {\it nonlocal relation} between displacement components. 
	
	The load on the coating, of components $p$ and $q$ in equations 
	(\ref{equilibrium})$_{1,2}$, 
(\ref{azione})$_{3}$, and (\ref{differenzialona}), 
	is partially applied externally to the disk, but also  transmitted by the disk in terms of stress  components $\sigma_{rr}$ and $\sigma_{r\theta}$ (multiplied by the thickness of the disk $b$, which becomes unity in a plane strain problem). Therefore, the loads $p$ and $q$ present in equations 
	(\ref{equilibrium})$_{1,2}$, 
(\ref{azione})$_{3}$, and (\ref{differenzialona}), have now to be interpreted as 
	\begin{equation}
	\label{Pforce}
	q \longrightarrow q -b \sigma_{r\theta}, ~~~\mbox{and}~~~ 
	p \longrightarrow p+b\sigma_{rr}, 
	\end{equation}
	so that in this way $p$ and $q$ are identified as external loads applied to the coated disk. In a complex notation, the total load applied on the coating can be represented
	as 
	\begin{equation}
	\label{Pfiga}
	P = p+b\sigma_{rr}  
	+ i \, \left(q -b \sigma_{r\theta}\right). 
	\end{equation}
	
	In the elastic disk, the strain components, expressed in polar coordinates $r$ and $\theta$, are 
	\begin{equation}
	\label{defni}
	\epsilon_{rr} =u_{r,r}, ~~~ \epsilon_{\theta \theta} = \frac{u_r+u_{\theta,\theta}}{r}, ~~~ 
	\epsilon_{r\theta}= \frac{u_{r,\theta}-u_\theta}{2r}  + \frac{u_{\theta, r}}{2} , 
	\end{equation}
	so that, using Hooke's laws, the stress components become
	\begin{equation}
	\label{stre}
	\sigma_{rr} =\mu \frac{(1+k) \epsilon_{rr} +(3-k) \epsilon_{\theta\theta}}{k-1} , 
	~~~ 
	\sigma_{\theta \theta} =\mu \frac{(1+k) \epsilon_{\theta\theta} +(3-k) \epsilon_{rr}}{k-1} , 
	~~~ 
	\sigma_{r\theta}= 2\mu\epsilon_{r\theta}, 
	\end{equation}
	and vice-versa
	\begin{equation}
	\label{stra}
	\epsilon_{rr} = \frac{(1+k) \sigma_{rr} +(k-3) \sigma_{\theta\theta}}{8\mu} , 
	~~~ 
	\epsilon_{\theta \theta} =\frac{(1+k) \sigma_{\theta\theta} +(k-3) \sigma_{rr}}{8\mu} , 
	~~~ 
	\epsilon_{r\theta}= \frac{\sigma_{r\theta}}{2\mu}, 
	\end{equation}
	where
	\begin{equation}
	\kappa = 
	\left\{
	\begin{array}{ll}
	\displaystyle 3 - 4 \nu, & \mbox{ for plane strain,} \\ [5 mm]
	\displaystyle \frac{3-\nu}{1+\nu}, & \mbox{ for plane stress.}
	\end{array}
	\right.
	\end{equation}

	
	\subsection{Complex potential representation for the disk}
	
	The elastic disk of radius $R$, enclosed by the annular beam defining the coating, has a 
	smooth, non intersecting boundary, so that the tangential and normal directions can be used and respectively denoted by the 
	unit vectors  ${\bf e}_r$ and ${\bf e}_\theta$. Every point of the disk can be identified by a complex number $z=x_{1}+ix_{2}$, 
	so that points belonging to the boundary $L$ will be denoted as $\tau=R\,e^{i\,\theta}$. 
	The following notation is introduced 
	\begin{equation}
	\label{gtau}
	\begin{aligned}
	g\left(\tau\right)=\frac{R}{\tau} , &&\overline{g\left(\tau\right)}=\frac{R}{\overline{\tau}}=g^{-1}\left(\tau\right), &&	g^{\prime}\left(\tau\right)=-\frac{1}{R}\,g^{2}\left(\tau\right),
	\end{aligned}
	\end{equation}
	where the prime, $(\,)^\prime$, denotes differentiation with respect to the variable $\tau$, while a superscript bar indicates complex conjugate.
	
	In a complex variable formulation, complex combinations are introduced for displacements, boundary tractions, and external load. These combinations allow the governing equations for the problem (\ref{equilibrium}), (\ref{inextensibility}) and (\ref{azione}) to be rewritten in a way that the elastic fields are determined in terms of complex Fourier series involving unknown complex coefficients.
	The knowledge of the latter coefficients permits the evaluation of 
	displacements and stresses everywhere in the disk via Kolosov-Muskhelishvili complex potentials $\varphi\left(z\right)$ and $\psi\left(z\right)$ defined as 
	\cite{muskhelishvili2013some} 
	\begin{equation}
	\label{muskhelishvilifields}
	\begin{aligned}
	\left\{\begin{array}{lll}
	2\mu{u\left(z\right)}=\kappa\varphi\left(z\right)-z\overline{\varphi^{\prime}\left(z\right)}-\overline{\psi\left(z\right)} ,\\
	\sigma_{11}+\sigma_{22}=4\,\mathrm{Re}\!\left(\varphi^{\prime}\left(z\right)\right) , \\
	\sigma_{22}-\sigma_{11}+2i\sigma_{12}=2\left[\overline{z}\varphi^{\prime\prime}\left(z\right)+\psi^{\prime}\left(z\right)\right] ,
	\end{array}\right.
	\end{aligned} 
	\end{equation}
	where $\mathrm{Re}\!\left(\,\,\right)$, and in the following $\mathrm{Im}\!\left(\,\,\right)$, represent the real and the imaginary part of the enclosed quantity.
	
	The components of the strain tensor can be expressed via the inverse Hooke's law as
			\begin{equation}
		\label{muskhelishvilistrain}
		\begin{aligned}
		\left\{\begin{array}{lll}
		\displaystyle \varepsilon_{11}+\varepsilon_{22}=2 \frac{1-2\nu}{\mu} \mathrm{Re}\!\left(\varphi^{\prime}\left(z\right)\right), \\
		\,\\
		\displaystyle \varepsilon_{22}-\varepsilon_{11}+2i\varepsilon_{12}=\frac{1}{\mu}\left[\overline{z}\varphi^{\prime\prime}\left(z\right)+\psi^{\prime}\left(z\right)\right]. 
		\end{array}
		\right.
		\end{aligned}
		\end{equation}

	When dealing with a circular disk, the general expression for the complex potentials $\varphi\left(z\right)$ and $\psi\left(z\right)$ has been stated by Mogilevskaya et al. \cite{mogilevskaya2008multiple} as
	\begin{equation}
	\label{potentialdisc}
	\begin{aligned}
	&\varphi\left(z\right)=\frac{2\mu}{\kappa-1}\,\mathrm{Re}\!\left(A_{1}\right)\,g^{-1}\left(z\right)+\frac{2\mu}{\kappa}\sum_{n=1}^{\infty}{A_{n+1}\,g^{-\left(n+1\right)}\left(z\right)} , \\
	&\psi\left(z\right)=-\frac{2\mu}{\kappa-1}\,\mathrm{Re}\!\left(A_{1}\right)\,\frac{\overline{z_{c}}}{R}-\frac{2\mu}{\kappa}\left[\frac{\overline{z_{c}}}{R}+g\left(z\right)\right]\sum_{n=1}^{\infty}{\left(n+1\right)A_{n+1}\,g^{-n}\left(z\right)}\\
	&\phantom{\psi\left(z\right)=}-2\mu\sum_{n=2}^{\infty}{\overline{A_{1-n}}\,g^{-\left(n-1\right)}\left(z\right)} ,
	\end{aligned}
	\end{equation}
	where $z_{c}$ denotes the centre of the disk and the functions $g\left(z\right),g^{\prime}\left(z\right)$ ,$g^{\prime\prime}\left(z\right)$, and the conjugate
	$\overline{g\left(z\right)}$ are defined as 
	\begin{equation}
	\label{gz}
	\begin{aligned}
	&g\left(z\right)=\frac{R}{z}=\frac{R}{\left(x_{1}+ix_{2}\right)} , \\[5pt]
	&g^{\prime}\left(z\right)=-\frac{1}{R}\,g^{2}\left(z\right) , \\[5pt]
	&g^{\prime\prime}\left(z\right)=\frac{2}{R^{2}}\,g^{3}\left(z\right) , \\[5pt]
	&\overline{g\left(z\right)}=\frac{R^{2}}{r^{2}}\,g^{-1}\left(z\right)  , \\[5pt]
	&r=\sqrt{x_{1}^{2}+x_{2}^{2}} \, .
	\end{aligned}
	\end{equation} 
	
	It will be shown that some of the coefficients $A_j$ ($j=1,...,+\infty$ and $j=-1,...,-\infty$) have to be evaluated by imposing additional conditions. In particular, the rigid body motion will be eliminated by fixing points belonging to $L$. When the latter condition is imposed, the expression for the displacement field in equation (\ref{muskhelishvilifields}) becomes 
	\begin{equation}
	\label{displrigidfixed}
		u\left(z\right)=\frac{1}{2\mu}\left[\kappa\varphi\left(z\right)-z\overline{\varphi^{\prime}\left(z\right)}-\overline{\psi\left(z\right)}\right]+A_{0}+i\,z\,\mathrm{Im}\!\left(A_{1}\right) .
	\end{equation}

	\section{Solution for the coated disk \label{coatdiskSol}}

	In a Cartesian coordinate system, the displacement at any point of the boundary of the disk, which coincides with the coating, $\tau\in{L}$ can be expressed through the complex Fourier series expansion 
	\begin{equation}
	\label{displacement}
		u\left(\tau\right)=u_{1}\left(\tau\right)+i\,u_{2}\left(\tau\right)=\sum_{n=1}^{\infty}{A_{-n}\,g^{n}\left(\tau\right)}+\sum_{n=0}^{\infty}{A_{n}\,g^{-n}\left(\tau\right)} ,
	\end{equation}
	where $u_{1}\left(\tau\right)$ and $u_{2}\left(\tau\right)$ are displacement components respectively parallel to the $x_1$ and $x_2$ axes and the complex coefficients $A_{\pm{n}}$ are 
	for the moment unknown.
	
	Recalling the relation between Cartesian and polar coordinates,
	\begin{equation}
		\label{transfrules}
		u_{r}\left(\tau\right)+i\,u_{\theta}\left(\tau\right)=\left[u_{1}\left(\tau\right)+i\,u_{2}\left(\tau\right)\right]g\left(\tau\right),
	\end{equation}
	it is possible to obtain the displacement at every point $\tau$ in the polar coordinate system $\left(r;\theta\right)$ as 
	\begin{equation}
		\label{ur}
		\begin{aligned}
			&u_{r}\left(\tau\right)=\frac{1}{2}\left[u\left(\tau\right)\,g\left(\tau\right)+\overline{u\left(\tau\right)}\,g^{-1}\left(\tau\right)\right] , \\
			&u_{\theta}\left(\tau\right)=\frac{1}{2i}\left[u\left(\tau\right)\,g\left(\tau\right)-\overline{u\left(\tau\right)}\,g^{-1}\left(\tau\right)\right] , 
		\end{aligned}
	\end{equation}
	so that the final representation for displacements follows from the equation (\ref{displacement})
	\begin{equation}
		\label{urComp}
		\begin{aligned}
			&u_{r}\left(\tau\right)=\frac{1}{2}\left[\sum_{n=1}^{\infty}{A_{-n}\,g^{n+1}\left(\tau\right)}+\sum_{n=0}^{\infty}{A_{n}\,g^{-\left(n-1\right)}\left(\tau\right)}\right.+\left.\sum_{n=1}^{\infty}{\overline{A_{-n}}\,g^{-\left(n+1\right)}\left(\tau\right)}+\sum_{n=0}^{\infty}{\overline{A_{n}}\,g^{n-1}\left(\tau\right)}\right] ,\\[5pt]
			&u_{\theta}\left(\tau\right)=\frac{1}{2i}\left[\sum_{n=1}^{\infty}{A_{-n}\,g^{n+1}\left(\tau\right)}+\sum_{n=0}^{\infty}{A_{n}\,g^{-\left(n-1\right)}\left(\tau\right)}\right.-\left.\sum_{n=1}^{\infty}{\overline{A_{-n}}\,g^{-\left(n+1\right)}\left(\tau\right)}-\sum_{n=0}^{\infty}{\overline{A_{n}}\,g^{n-1}\left(\tau\right)}\right] . \\
		\end{aligned}
	\end{equation} 
	
	
	In a similar vein, the tractions at any point of the coating $\tau$ can be expressed through the complex Fourier series 
	\begin{equation} 
	\label{traction}
	\sigma\left(\tau\right)=\sigma_{rr}\left(\tau\right)+i\,\sigma_{r\theta}\left(\tau\right)=\sum_{n=1}^{\infty}{B_{-n}\,g^{n}\left(\tau\right)}+\sum_{n=0}^{\infty}{B_{n}\,g^{-n}\left(\tau\right)} ,
	\end{equation}
	where $\sigma_{rr}$ and $\sigma_{r\theta}$ are the components of the traction at the point $\tau\in{L}$ 
respectively	directed parallel to the normal and the tangential direction.

	The coefficients $A_{\pm{n}}$ and $B_{\pm{n}}$ are interrelated through the following relations (see Zemlyanova and Mogilevskaya \cite{zemlyanova2018circular}) 
	\begin{equation}
	\label{ABinterrelation}
	\begin{aligned}
	&B_{-1}=0 , \\
	&\frac{\kappa-1}{2\mu}\,B_{0}=\frac{2}{R}\,\mathrm{Re}\!\left(A_{1}\right) , \\
	&\frac{1}{2\mu}\,B_{-n}=\frac{n-1}{R}\,A_{1-n} , \,&\text{for}&&n\ge{2} , \\
	&\frac{\kappa}{2\mu}\,B_{n}=\frac{n+1}{R}\,A_{n+1} , \,&\text{for}&&n\ge{1} .
	\end{aligned}
	\end{equation}

	The representation of the load acting on the external surface of the coating  is introduced  in its local coordinates system, 
	using the following   
complex Fourier series as (Fig. \ref{coat22}):
	\begin{equation}
		\label{p+q*}
			p+iq=\sum_{n=1}^{\infty}{D_{-n}}\,g^{n}\left(\tau\right)+\sum_{n=0}^{\infty}{D_{n}\,g^{-n}\left(\tau\right)},
	\end{equation}
	where the components of the load $p$ and $q$ are assumed to be single-valued on $L$ and variable with the angle $\theta$, while  coefficients 
	$D_{\pm{n}}$ 
	remain determined as the series expansion of $p$ and $q$, so that they are treated as known 
	complex coefficients. The complex representation of the total load affecting the coating,  $P$, equation (\ref{Pfiga}), can be computed by recalling equations \eqref{traction} and \eqref{p+q*} to obtain 
	\begin{equation}
	\label{Ptau}
	\begin{aligned}
		P\left(\tau\right)=\sum_{n=1}^{\infty}{\left(D_{-n}+b\,\overline{B_{n}}\right)\,g^{n}\left(\tau\right)}+\sum_{n=0}^{\infty}{\left(D_{n}+b\,\overline{B_{-n}}\right)\,g^{-n}\left(\tau\right)} .
	\end{aligned}
	\end{equation}

		\subsection{\small{Modelling of inextensible coating}}
	
	The axial strain in the coating, equation (\ref{cinema})$_1$, can be translated into the complex notation as 
	shown in \cite{zemlyanova2018circular}, their equation (99)$_2$, namely,  
	\begin{equation}
	\label{ZM2018}
	\begin{aligned}
		u_{\theta,\theta}\left(\tau\right)+u_{r}\left(\tau\right)=R\,\mathrm{Re}\!\left(\frac{\partial{u}\left(\tau\right)}{\partial{\tau}}\right) ,
	\end{aligned}
	\end{equation}
	so that the inextensibility condition for the annular beam, in other words the isoperimetric constraint, equation \eqref{inextensibility}, becomes 
		\begin{equation}
		\label{inextensibility1}
			\mathrm{Re}\!\left(\frac{\partial{u\left(\tau\right)}}{\partial{\tau}}\right)=0 . 
	\end{equation}
	Differentiating equation \eqref{displacement} with respect to $\tau$ and using the resulting $g^\prime$ into equation (\ref{gtau})$_{3}$ yields
	\begin{equation}
	\label{dudtau}
	\begin{aligned}
		\frac{\partial{u\left(\tau\right)}}{\partial{\tau}}=\frac{1}{R}\sum_{n=1}^{\infty}{\left(-n\,A_{-n}\,g^{n+1}\left(\tau\right)+n\,A_{n}\,g^{-\left(n-1\right)}\left(\tau\right)\right)} ,
	\end{aligned}
	\end{equation}
	so that, a substitution into equation \eqref{inextensibility1} leads to
	\begin{equation}
	\label{re=0}
	\sum_{n=1}^{\infty}\left(n\,A_{-n}\,g^{n+1}\left(\tau\right)-n\,A_{n}\,g^{-\left(n-1\right)}\left(\tau\right) +
	n\,\overline{A_{-n}}\,g^{-\left(n+1\right)}\left(\tau\right)-n\,\overline{A_{n}}\,g^{n-1}\left(\tau\right)\right)=0 .
	\end{equation}
	By collecting terms with the same power of $g^{\pm{n}}\left(z\right)$ in the left hand side of equation \eqref{re=0} and by equating them to zero leads to the following conditions.
	\begin{enumerate}[label=(\roman*.)]
		\item For $n=0$:
			\begin{equation}
			\label{A1}
			\mathrm{Re}\left(A_{1}\right)=0 ,
			\end{equation}
	
		\item For $n=-1$:
			\begin{equation}
			\label{A2}
			\begin{aligned}
				A_{2}=0 ,
			\end{aligned}
			\end{equation}
			
			
		\item For $n\neq{0}$ and $n\neq{-1}$:
		\begin{equation}
		\label{Ak=A-kconj}
			\begin{aligned}
				A_{n+1}=\frac{n-1}{n+1}\,\overline{A_{1-n}} .
			\end{aligned}
		\end{equation}
	\end{enumerate}
	Recalling the relation between coefficients $A_{n}$ and $B_{n}$ expressed by equations (\ref{ABinterrelation}), $B_{0}=0$ follows  from equation \eqref{A1}. 
	Moreover, setting $n=1$ in equation \eqref{ABinterrelation}$_{4}$ and recalling equation \eqref{A2}, $B_{1}=0$ follows.

	\subsection{\small{Equilibrium and kinematic condition}}

	The 5-th order differential equation \eqref{differenzialona} describing the radial displacement couples with equilibrium and kinematics of the beam representing the coating, under the constraint of axial inextensibility \eqref{inextensibility}. The perfect bonding between coating and disk, equation \eqref{continuity} implies that the radial displacement entering equation \eqref{differenzialona} coincides with the radial displacement at the boundary of the disk. 
		In particular, it is possible to rewrite the differential equation \eqref{differenzialona} through an identification of the components of the external load 
		$p$ and $q$ provided by equation  \eqref{Pforce}. At the right hand side of equation \eqref{differenzialona}, the quantity $\left(q -d{p}/d\theta\right)$ assumes the form 
	\begin{equation}
	\label{rhsDifferenzialona}
		q- \frac{dp}{d\theta}=q-p_{,\theta}-b\left[\sigma_{r\theta}\left(\tau\right)+\sigma_{rr,\theta}\left(\tau\right)\right] ,
	\end{equation}
	where the terms on the right hand side can be particularized by exploiting the similarity with equation (98) in \cite{zemlyanova2018circular} as 
	\begin{equation}
		\label{qdp}
		\begin{aligned}
			&\frac{q-p_{,\theta}}{R}= \mathrm{Im}\!\left(\frac{\partial}{\partial{\tau}}\left[{\left(p+i\,q\right)g^{-1}\left(\tau\right)}\right]\right) ,\\
			&\sigma_{r\theta}\left(\tau\right)+\sigma_{rr,\theta}\left(\tau\right)=2\sigma_{r\theta}\left(\tau\right)-R\,\mathrm{Im}\!\left(\frac{\partial}{\partial{\tau}}{\left[\sigma\left(\tau\right)g^{-1}\left(\tau\right)\right]}\right) .
	\end{aligned}
	\end{equation}
	
	Equation \eqref{gz}$_{2}$ for the derivatives and expressions \eqref{traction} and \eqref{p+q*} lead to 
	\begin{equation}
	\label{qdpexpr}
	\begin{aligned}
		q-p_{,\theta}=&-\frac{1}{2i}\left\{\sum_{n=1}^{\infty}{\left[\left(n-1\right)D_{-n}+\left(n+1\right)\overline{D_{n}}\,\right]g^{n}\left(\tau\right)}\right. \\
		&\left.-\sum_{n=1}^{\infty}{\left[\left(n+1\right)D_{n}+\left(n-1\right)\overline{D_{-n}}\,\right]g^{-n}\left(\tau\right)}\right\} +\mathrm{Im}\!\left(D_{0}\right) ,\\
		\sigma_{r\theta}\left(\tau\right)+\sigma_{rr,\theta}\left(\tau\right)=&\frac{1}{2i}\left\{\sum_{n=1}^{\infty}{\left[\left(n+1\right)B_{-n}+\left(n-1\right)\overline{B_{n}}\,\right]g^{n}\left(\tau\right)}\right. \\
		&\left.-\sum_{n=1}^{\infty}{\left[\left(n-1\right)B_{n}+\left(n+1\right)\overline{B_{-n}}\,\right]g^{-n}\left(\tau\right)}\right\} +\mathrm{Im}\!\left(B_{0}\right) .
	\end{aligned}
	\end{equation}
	
	Through a substitution of equations \eqref{qdpexpr}, the expression \eqref{rhsDifferenzialona} can now be cast in the form
	\begin{equation} 
	\label{rhsDifferenzialonaexpr}
	\begin{aligned}
		q-\frac{dp}{d\theta}&=-\frac{1}{2i}\left\{\sum_{n=1}^{\infty}{\left[\left(n+1\right)\left(\overline{D_{n}}+b\,B_{-n}\right)+\left(n-1\right)\left(D_{-n}+b\,\overline{B_{n}}\right)\right]g^{n}\left(\tau\right)}\right. \\
		&\left.-\sum_{n=1}^{\infty}{\left[\left(n+1\right)\left(D_{n}+b\,\overline{B_{-n}}\right)+\left(n-1\right)\left(\overline{D_{-n}}+b\,B_{n}\right)\right]g^{-n}\left(\tau\right)}\right\} +\mathrm{Im}\!\left(D_{0}-b\,B_{0}\right) .
	\end{aligned}
	\end{equation}
	
	The left hand side of equation \eqref{differenzialona} can be written in terms of coefficients $A_{\pm{n}}$  by computing the complex derivatives of the displacement in equation \eqref{ur}$_{2}$. In particular, the same procedure reported in \cite{zemlyanova2018circular}, their Appendix C, yields now
	\begin{equation} 
	\label{lhsdifferenzialona}
	\begin{aligned}
		\frac{d^5 u_r}{d\theta^5}+2 \frac{d^3 u_r}{d\theta^3} + \frac{d u_r}{d\theta}=&-R^{5}\,\mathrm{Im}\!\left(\frac{\partial^{5}{u\left(\tau\right)}}{\partial{\tau^{5}}}g^{-4}\left(\tau\right)\right)-5R^{4}\,\mathrm{Im}\!\left(\frac{\partial^{4}{u\left(\tau\right)}}{\partial{\tau^{4}}}g^{-3}\left(\tau\right)\right)\\
		&-3R^{3}\,\mathrm{Im}\!\left(\frac{\partial^{3}{u\left(\tau\right)}}{\partial{\tau^{3}}}g^{-2}\left(\tau\right)\right) ,
	\end{aligned}
	\end{equation}
	where, in agreement with \cite{zemlyanova2018circular} [their equation (100)], the derivatives of the complex representation of the displacement field reads as
	\begin{equation}
	\label{urdkth}
	\begin{aligned}
		\frac{\partial^{3}u\left(\tau\right)}{\partial{\tau^{3}}}=&\frac{1}{R^{3}}\left[-\sum_{n=1}^{\infty}{n\left(n+1\right)\left(n+2\right)A_{-n}\,g^{n+3}\left(\tau\right)}\right.\\
		&\left.+\sum_{n=3}^{\infty}{n\left(n-1\right)\left(n-2\right)A_{n}\,g^{-\left(n-3\right)}\left(\tau\right)}\right] ,\\
		\frac{\partial^{4}u\left(\tau\right)}{\partial{\tau^{4}}}=&\frac{1}{R^{4}}\left[+\sum_{n=1}^{\infty}{n\left(n+1\right)\left(n+2\right)\left(n+3\right)A_{-n}\,g^{n+4}\left(\tau\right)}\right.\\
		&\left.+\sum_{n=4}^{\infty}{n\left(n-1\right)\left(n-2\right)\left(n-3\right)A_{n}\,g^{-\left(n-4\right)}\left(\tau\right)}\right] ,\\
		\frac{\partial^{5}u\left(\tau\right)}{\partial{\tau^{5}}}=&\frac{1}{R^{5}}\left[-\sum_{n=1}^{\infty}{n\left(n+1\right)\left(n+2\right)\left(n+3\right)\left(n+4\right)A_{-n}\,g^{n+5}\left(\tau\right)}\right.\\
		&\left.+\sum_{n=5}^{\infty}{n\left(n-1\right)\left(n-2\right)\left(n-3\right)\left(n-4\right)A_{n}\,g^{-\left(n-5\right)}\left(\tau\right)}\right] .
	\end{aligned}
	\end{equation}
	
	A complete form of equation \eqref{lhsdifferenzialona} can be computed after substitution of expressions \eqref{urdkth} as 
	\begin{equation}
	\label{lhsdifferenzialonaexpr}
	\begin{aligned}
			\frac{d^5 u_r}{d\theta^5}+2 \frac{d^3 u_r}{d\theta^3} + \frac{d u_r}{d\theta}=&\frac{1}{2i}\left\{\sum_{n=1}^{\infty}{n^{2}\left(n+1\right)\left(n+2\right)^{2}\left[A_{-n}\,g^{n+1}\left(\tau\right)-\overline{A_{-n}}\,g^{-\left(n+1\right)}\left(\tau\right)\right]}\right.\\
			&\left.-\sum_{n=3}^{\infty}{n^2\left(n-1\right)\left(n-2\right)^{2}\left[A_{n}\,g^{-\left(n-1\right)}\left(\tau\right)-\overline{A_{n}}\,g^{n-1}\left(\tau\right)\right]}\right\} .
	\end{aligned}
	\end{equation}
	
	Using expressions \eqref{rhsDifferenzialonaexpr} and \eqref{lhsdifferenzialonaexpr} and collecting terms with the same summation index, the differential equation \eqref{differenzialona} becomes 
	\begin{equation} 
	\label{cmpdifferenzialona}
	\begin{aligned}
		&-\sum_{n=1}^{\infty}{\frac{R^{4}}{EJ}\left\{\left[\left(n+1\right)\left(\overline{D_{n}}+b\,B_{-n}\right)+\left(n-1\right)\left(D_{-n}+b\,\overline{B_{n}}\right)\right]g^{n}\left(\tau\right)\right.}\\[3 mm]
		&\left.-\left[\left(n+1\right)\left(D_{n}+b\,\overline{B_{-n}}\right)+\left(n-1\right)\left(\overline{D_{-n}}+b\,B_{n}\right)\right]g^{-n}\left(\tau\right)-2i\,\mathrm{Im}\!\left(D_{0}-b\,B_{0}\right)\right\}\\[3 mm]
		&-n^{2}\left(n+1\right)\left(n+2\right)^{2}\left[A_{-n}\,g^{n+1}\left(\tau\right)-\overline{A_{-n}}\,g^{-\left(n+1\right)}\left(\tau\right)\right]\\[3 mm]
		&+\sum_{n=3}^{\infty}{n^{2}\left(n-1\right)\left(n-2\right)^{2}\left[A_{n}\,g^{-\left(n-1\right)}\left(\tau\right)-\overline{A_{n}}\,g^{n-1}\left(\tau\right)\right]}=0 ,
	\end{aligned}
	\end{equation}
	from which, collecting terms with the same power $g^{\pm{n}}\left(\tau\right)$ in equation \eqref{cmpdifferenzialona} and equating them to zero, the following expression 
	is obtained
		\begin{equation} 
	\label{A-k+1}
	\begin{array}{ll}
		\left(n+1\right)\left(\overline{D_{n}}+b\,B_{-n}\right)+\left(n-1\right)\left(D_{-n}+b\,\overline{B_{n}}\right) \\ [5 mm]
			~~~~~ ~~~~~ ~~~~~ 
			\displaystyle 
			+\frac{n\left(n+1\right)^{2}\left(n-1\right)^{2} EJ}{R^4}\left[A_{1-n}+\overline{A_{n+1}}\,\right]=0 ,
	\end{array}
	\end{equation}
	of which the following two particular cases can be highlighted
	\begin{equation}
	\label{D=-D-1}
	\mathrm{Im}\!\left(D_{0}-bB_{0}\right)=0 , ~~\mbox{ for } ~ n=0 
	~~\mbox{ and } ~~ 
	\overline{B_{-1}}=-\frac{1}{b}D_{1}, 
	~~\mbox{ for } ~ n=-1.
	\end{equation}

	Recalling equations \eqref{ABinterrelation}$_{3-4}$ and \eqref{Ak=A-kconj}, it becomes possible to express the coefficients $B_{\pm{n}}$ involved in equation \eqref{A-k+1} as functions of the coefficients $A_{\pm{n}}$ 
	\begin{equation}
	\label{BfncA}
	\begin{aligned}
		&\overline{B_{n}}=\frac{2\mu}{\kappa{R}}\left(n+1\right)\overline{A_{n+1}}=\frac{2\mu}{\kappa{R}}\left(n-1\right)A_{1-n} ,\,&&\text{for}\,n\ge{1}\\
		&B_{-n}=\frac{2\mu}{R}\left(n-1\right)A_{1-n} ,\,&&\text{for}\,n\ge{2} .
	\end{aligned}
	\end{equation}
	
	A substitution of equations \eqref{BfncA} into equation \eqref{A-k+1} yields 
	\begin{equation}
	\label{A-k+1(D)}
	\begin{aligned}
		& \left(n+1\right)\overline{D_{n}}+\frac{2\mu{b}}{R}\left(n^{2}-1\right)A_{1-n}+\left(n-1\right)D_{-n}+\frac{2\mu{b}}{\kappa{R}}\left(n-1\right)^{2}A_{1-n} \\[5pt]
		& ~~~~~ ~~~~~ +\frac{2n^{2}\left(n+1\right)\left(n-1\right)^{2} EJ}{R^4} A_{1-n}=0 ,
	\end{aligned}
	\end{equation}
	so that collecting terms involving $A_{1-n}$ leads to
	\begin{equation}
	\label{A-k+1(Df)} 
	\begin{aligned}
		A_{1-n}=-\frac{\kappa{R^{4}}\left[\left(n+1\right)\overline{D_{n}}+\left(n-1\right)D_{-n}\right]}{2\left(n-1\right)\left[\kappa{EJ}n^{2}\left(n^{2}-1\right)+R^{3}\mu{b}\left(n+\kappa{n}+\kappa-1\right)\right]} ,\,&&\text{for}\,n\ge{2} .
	\end{aligned}
	\end{equation}


	\subsection{\small{Fixing rigid body motion}}
	
	Equation \eqref{displrigidfixed} shows that restrictions on coefficients $A_{0}$ and $A_{1}$ can be inferred from the imposition of a rigid-body roto-translation. Following the procedure adopted in \cite{mogilevskaya2008multiple}, the latter will be eliminated by assuming the displacement of the disk to be zero at one point of it and requiring vanishing of the vertical displacement component at another point of the disk on the same horizontal line. 
		It will now be imposed $u\left(z\right)=0$ at points $z=\tau_{0}=R+i\,0$ and $z=z_{c}=0$, see Fig. \eqref{coat22} a. 
	Recalling equations \eqref{A1}-\eqref{Ak=A-kconj} and setting $z_{c}=0$ in relations \eqref{potentialdisc}, the complex potentials for the disk are determined as 
	\begin{equation}
	\label{potentialdisczc}
	\begin{aligned}
		&\varphi\left(z\right)=\frac{2\mu}{\kappa}\sum_{n=1}^{\infty}{A_{n+1}\,g^{-\left(n+1\right)}\left(z\right)} ,\\
		&\psi\left(z\right)=-2\mu\sum_{n=2}^{\infty}{\left(\frac{n+\kappa-1}{\kappa}\right)\overline{A_{1-n}}\,g^{-\left(n-1\right)}\left(z\right)} .\\
	\end{aligned}
	\end{equation}
	At points $z=z_{c}$  and $z=\tau_{0}$ the following results can be derived from equations \eqref{displrigidfixed}. 
	
	\begin{itemize}
	\item At $z=z_{c}$: 
	\begin{equation}
	\label{u0}
	\begin{aligned}
		{u}\left(z_{c}\right)=\frac{1}{2\mu}\left[\kappa\varphi\left(z_{c}\right)-\overline{\psi\left(z_{c}\right)}\right]+A_{0} ,
	\end{aligned}
	\end{equation}
	so that 
	equations \eqref{gz}$_{1-2}$, \eqref{A1} and \eqref{A2} lead to the particularization of equation \eqref{potentialdisczc} at $z=z_{c}$ 
	\begin{equation}
	\label{potentialdiscz}
	\begin{aligned}
		\varphi\left(z_{c}\right)=\psi\left(z_{c}\right)=0 , 
	\end{aligned}
	\end{equation}
	in addition to which equations \eqref{potentialdiscz}, substituted into equation \eqref{u0} with $u\left(z_{c}\right)=0$, provide 
	$A_{0}=0$.

		\item At $z=\tau_{0}$:
	\begin{equation}
	\label{utaufx}
	\begin{aligned}
		{u}\left(\tau_{0}\right)=\frac{1}{2\mu}\left[\kappa\varphi\left(\tau_{0}\right)-\tau_{0}\,\overline{\varphi^{\prime}\left(\tau_{0}\right)}-\overline{\psi\left(\tau_{0}\right)}\right]+i\,\tau_{0}\,\mathrm{Im}\!\left(A_{1}\right) ,
	\end{aligned}
	\end{equation}
	from which the following specific expressions for the complex potentials \eqref{potentialdisczc} can be derived 
	\begin{equation}
	\label{potentialzcR}
	\begin{aligned}
		&\varphi\left(\tau_{0}\right)=\frac{2\mu}{\kappa}\sum_{n=1}^{\infty}{A_{n+1}} , \\[10pt]
		&\varphi^{\prime}\left(\tau_{0}\right)=\frac{2\mu}{R\kappa}\sum_{n=1}^{\infty}{\left(n+1\right)}A_{n+1} ,
	\end{aligned}
	\end{equation}
	and 
	\begin{equation}
	\label{potentialzcR2}
	\begin{aligned}
		\psi\left(\tau_{0}\right)=-2\mu\sum_{n=2}^{\infty}{\left(\frac{n+\kappa-1}{\kappa}\right)\overline{A_{1-n}}} ,
	\end{aligned}
	\end{equation}
	where the latter expressions have been obtained by setting $\tau_{0}=R$ in equations \eqref{gtau}, i.e $g\left(\tau_{0}\right)^{-n}=\left(\tau_{0}/R\right)^{n}=1\,\,\forall{n\geq{0}}$. An explicit expression for $u\left(\tau_{0}\right)$ can be obtained through a substitution of equations \eqref{potentialzcR} and \eqref{potentialzcR2} into equation 
	\eqref{utaufx} 
	\begin{equation} 
	\label{u(R)=0}
	\begin{aligned}
		{u}\left(\tau_{0}\right)&=\sum_{n=1}^{\infty}{A_{n+1}}-\frac{1}{\kappa}\sum_{n=1}^{\infty}{\left(n+1\right)\overline{A_{n+1}}}+\sum_{n=2}^{\infty}{\left(\frac{n+\kappa-1}{\kappa}\right)A_{1-n}}\\
	& +iR\,\,\mathrm{Im}\!\left(A_{1}\right) ,
	\end{aligned}
	\end{equation}
	which, recalling equations \eqref{A2} and \eqref{Ak=A-kconj}, provides 
	\begin{equation}
	\label{utau0}
		u\left(\tau_{0}\right)=\sum_{n=2}^{\infty}{\left[\frac{n-1}{n+1}\,\overline{A_{1-n}}+A_{1-n}\right]+iR\,\mathrm{Im}\!\left(A_{1}\right)} .
	\end{equation}
	Imposition of  the condition $u_{2}\left(\tau_{0}\right)=\mathrm{Im}\!\left(u\left(\tau_{0}\right)\right)$ in equation \eqref{utau0} yields 
	\begin{equation} 
	\label{Imutau0}
		u_{2}\left(\tau_{0}\right)=\mathrm{Im}\!\left(\sum_{n=2}^{\infty}{\left[\frac{n-1}{n+1}\,\overline{A_{1-n}}+A_{1-n}\right]}\right)+R\,\mathrm{Im}\!\left(A_{1}\right) .
	\end{equation}
	The requirement that the vertical displacement component $u_{2}\left(\tau_{0}\right)$ be zero at 
	$z=\tau_{0}$ 
	is finally established from the vanishing of the right hand side of equation \eqref{Imutau0}, 
	\begin{equation}
	\label{ImA1}
		\mathrm{Im}\!\left(A_{1}\right)=-\frac{2}{R}\,\mathrm{Im}\!\left(\sum_{n=2}^{\infty}{\frac{1}{n+1}\,A_{1-n}}\right) .
	\end{equation}

		\end{itemize}

	The condition $A_0=0$ at $z=z_c$ and equation \eqref{ImA1} allow to derive all additional terms in equation \eqref{displrigidfixed}, describing a rigid body movement.

	The above obtained solution is summarized in terms of the evaluated coefficients reported in Tables \ref{TabA} and \ref{TabB}. These coefficients completely define the displacement and stress fields for the coated disk.
		\begin{center}
		\begin{minipage}{\linewidth}
			\centering
			\begin{tabular}[h]{c@{\hspace{4\tabcolsep}}c@{\hspace{4\tabcolsep}}c@{\hspace{4\tabcolsep}}}\toprule[1.5pt]
				\bf Coefficient & \bf Value & \bf Rule\\\midrule
				$A_{1-n}$& $-\frac{\kappa{R^{4}}\left[\left(n+1\right)\overline{D_{n}}+\left(n-1\right)D_{-n}\right]}{2\left(n-1\right)\left[\kappa{EJ}n^{2}\left(n^{2}-1\right)+R^{3}\mu{b}\left(n+\kappa{n}+\kappa-1\right)\right]}$& $n\geq{2}$\\[10pt]
				$A_{0}=A_{2}$& $0$  &-\\[10pt]
				$A_{1}$ & $-\frac{2\,i}{R}\,\mathrm{Im}\!\left(\sum_{n=2}^{\infty}{\frac{1}{n+1}\,A_{1-n}}\right)$ &-\\[10pt]
				$A_{n+1}$& $\frac{n-1}{n+1}\,\overline{A_{1-n}}$ & $n$>${1}$\\[10pt]
				\bottomrule
			\end{tabular}\par
			\captionof{table}{\label{TabA}{Coefficients $A_{\pm{n}}$ of the complex power series defining the displacement field within the disk.}}
		\end{minipage}
	\end{center}
	
Coefficients $A_{\pm{n}}$ listed in Table \ref{TabA} are given in terms of the known complex coefficients $D_{\pm{n}}$, which define the  loads applied to the external surface of the coated disk.

	\begin{center}
		\begin{minipage}{\linewidth}
			\centering
			\begin{tabular}[h]{c@{\hspace{4\tabcolsep}}c@{\hspace{4\tabcolsep}}c@{\hspace{4\tabcolsep}}}\toprule[1.5pt]
				\bf Coefficient & \bf Value & \bf Rule\\\midrule
				$B_{-n}$& $\frac{2\mu}{R}\left(n-1\right)A_{1-n}$ &$k\ge{2}$\\[10pt]
				$B_{-1}=B_{0}=B_{1}$& $0$ &-\\[10pt]
				$B_{n}$& $\frac{2\mu}{\kappa{R}}\left(n-1\right)\overline{A_{1-n}}$ &$k\ge{1}$\\[10pt]
				\bottomrule
			\end{tabular}\par
			\captionof{table}{\label{TabB}{Coefficients $B_{\pm{n}}$ of the complex power series defining the stress distribution inside the disk.}}
		\end{minipage}
	\end{center}
Table \ref{TabB} shows that all coefficients $B_{\pm{n}}$ can be written in terms of coefficients $A_{1-n}$ and hence from equation \eqref{A-k+1(Df)} they are found to depend on the known complex coefficients $D_{\pm{n}}$. Once $A_{\pm{n}}$ and $B_{\pm{n}}$ are known, it becomes possible to compute the expressions for elastic displacement and stress fields.
	
	\subsection{\small{Elastic fields for the disk}}	
	Displacement, stress, and strain at any point inside the disk and on its boundary are known in terms of 
	complex potentials $\varphi\left(z\right)$ and $\psi\left(z\right)$, equations \eqref{muskhelishvilifields}, through coefficients $A_{\pm{n}}$ listed in Table \ref{TabA}, as
	\begin{equation} 
	\label{potentialVarphi}
	\begin{aligned}
		&\varphi\left(z\right)=\frac{2\mu}{\kappa}\sum_{n=1}^{\infty}{\frac{n-1}{n+1}\,\overline{A_{1-n}}}\,g^{-\left(n+1\right)}\left(z\right) ,\\
		&\varphi^{\prime}\left(z\right)=\frac{2\mu}{R\kappa}\sum_{n=1}^{\infty}{\left(n-1\right)\overline{A_{1-n}}\,g^{-n}\left(z\right)} ,\\
		&\varphi^{\prime\prime}\left(z\right)=\frac{2\mu}{R^{2}\kappa}\sum_{n=1}^{\infty}{n\left(n-1\right)\overline{A_{1-n}}\,g^{-n+1}\left(z\right)} ,
	\end{aligned}
	\end{equation}
	\begin{equation} 
	\label{potentialPsi}
	\begin{aligned}
		&\psi\left(z\right)=-2\mu\sum_{n=2}^{\infty}{\left(\frac{n+\kappa-1}{\kappa}\right)\overline{A_{1-n}}\,g^{-n+1}\left(z\right)} ,\\
		&\psi^{\prime}\left(z\right)=-\frac{2\mu}{R}\sum_{n=2}^{\infty}{\left(n-1\right)\left(\frac{n+\kappa-1}{\kappa}\right)\,\overline{A_{1-n}}\,g^{-n+2}\left(z\right)} .
	\end{aligned}
	\end{equation}
	Except for a rigid roto-translation, the displacement $u\left(z\right)$ at every point of the disk is determined by relation \eqref{muskhelishvilifields}$_{1}$, through 
	equation \eqref{displrigidfixed} and the coefficients listed in Tables \ref{TabA} and \ref{TabB} entering the following equation
	\begin{equation}
	\label{uz}
	\begin{aligned}
		&u\left(z\right)=\sum_{n=1}^{\infty}{A_{n+1}g^{-n-1}\left(z\right)}+\sum_{n=2}^{\infty}{A_{1-n}g^{n-1}\left(z\right)}-2i\,g^{-1}\left(z\right)\,\mathrm{Im}\!\left(\sum_{n=2}^{\infty}{\frac{1}{n+1}\,A_{1-n}}\right) ,
	\end{aligned}
	\end{equation}
	which, recalling equations \eqref{A2} and \eqref{Ak=A-kconj} and collecting terms, can be rewritten as
	\begin{equation}
	\label{uzf}
	\begin{aligned}
		u\left(z\right)=&\sum_{n=2}^{\infty}{\left[\frac{n-1}{n+1}\overline{A_{1-n}}\,g^{-\left(n+1\right)}\left(z\right)+A_{1-n}\,g^{n-1}\left(z\right)\right.}\\
		&{\left.-2i\,g^{-1}\left(z\right)\mathrm{Im}\!\left(\frac{1}{n+1}\,A_{1-n}\right)\right]} .
	\end{aligned}
	\end{equation}

	
	The stress is determined from a combination between equations \eqref{muskhelishvilifields}$_{2-3}$, with a chain substitution of equations \eqref{A2} and \eqref{Ak=A-kconj} into the potentials, equations \eqref{potentialdisczc}$_{1}$, as 
	\begin{equation}
	\label{stresscombinationP}
	\begin{aligned}
		\sigma_{11}+\sigma_{22}=\frac{8\mu}{R\kappa}\mathrm{Re}\!\left(\sum_{n=2}^{\infty}{\left(n-1\right)\overline{A_{1-n}}\,g^{-n}\left(z\right)}\right) ,
	\end{aligned}
	\end{equation}
	while equation \eqref{muskhelishvilifields}$_{3}$ leads to 
	\begin{equation}
	\label{stresscombinationM}
	\begin{aligned}
		&\sigma_{22}-\sigma_{11}+2i\sigma_{12}=\frac{4\mu}{R}\sum_{n=2}^{\infty}{\left(n-1\right)\left[\frac{r^{2}}{R^{2}\kappa}n-\frac{1}{\kappa}\left(n-1\right)-1\right]\overline{A_{1-n}}\,g^{-n+2}\left(z\right)} .
	\end{aligned}
	\end{equation}
	
	\section{Experiments vs. theoretical predictions}
	
	The aim of this section is to introduce the experimental set-up and results and compare the latter with the general solution for the coated disk, to be particularized now to a specific applied loading,  tailored to model the experiments.
	
	\subsection{Photoelastic experiments \label{PhSec}}
	
\begin{figure}[ht!]
		\centering
		\subfigure{
			\includegraphics[keepaspectratio]{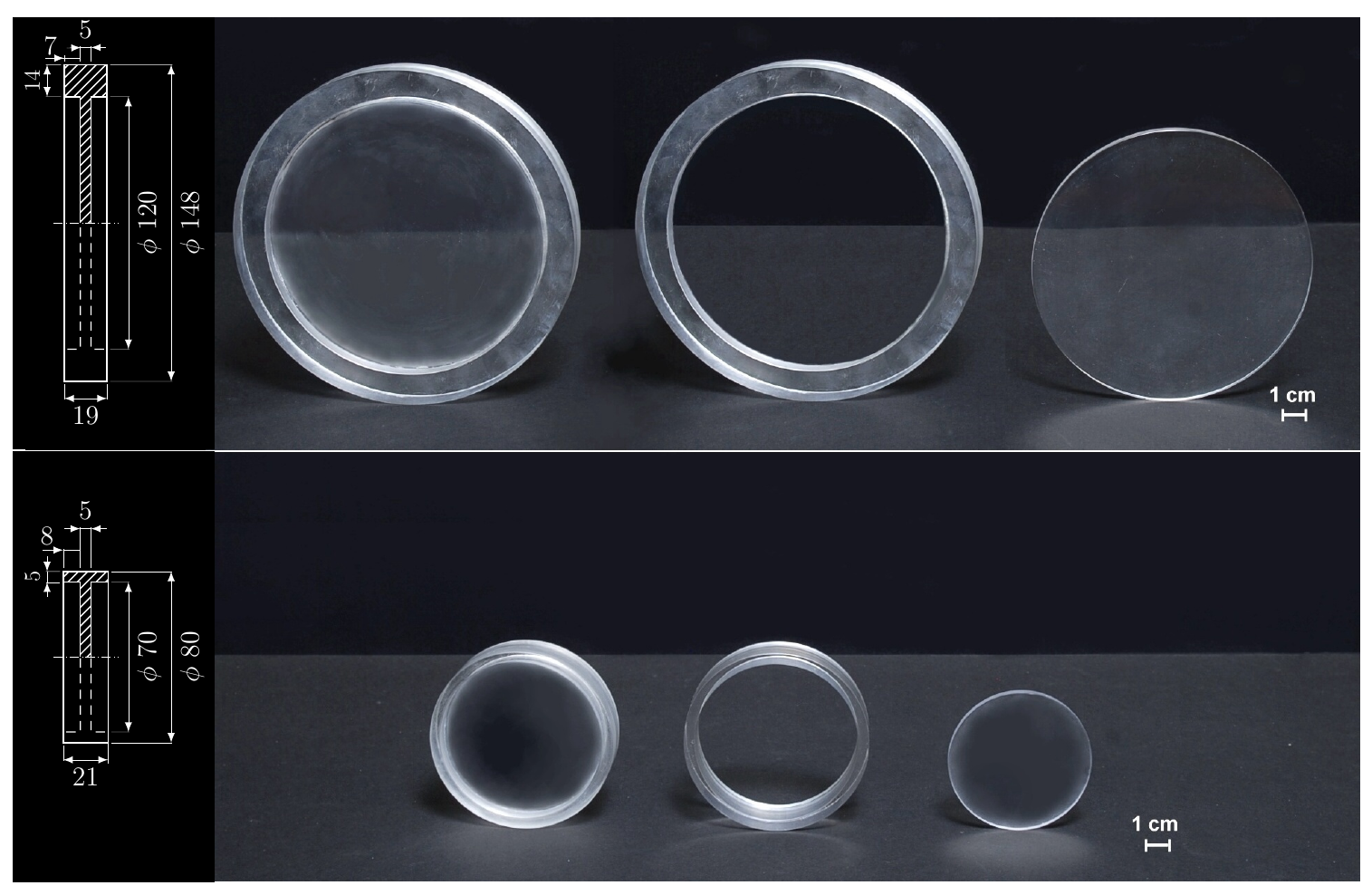}
		}
		\caption{The two families of samples used for diametrical compression tests: the coated disk (left), the coating alone, representing a circular beam  (center) and disk without coating (right). 
				The coating has a rectangular cross-section equal to $14 \times 19$~mm (upper part) and $5\times 21$~mm (lower part), resulting in a strongly different bending stiffness for the coating, when the two families of samples are compared.
		}
		\label{sample_Geom}
	\end{figure}

	The previously derived analytic solution is compared with photoelastic experiments performed at the Instabilities Lab of the University of Trento on two coated disks, subject to 
	two radial compressive force distributions applied at two diametrically opposed small portions  of the coating. 
	The two coated disks, shown in Fig. \ref{sample_Geom} on the left, have been 
	machined
	with radii $R_1$ and $R_{2}$ equal to 60 and 35~mm and thickness 5~mm in the internal portion (with a CNC engraving machine) from a single thick 
	plate of polymethyl methacrylate (Perspex Clear, Lucite International, 
	Young modulus $E=$~3200~MPa and Poisson's ratio $\nu = 0.36$). 
	The coatings have in-plane $\times$ out-of-plane thicknesses equal to 14~mm $\times$ 19~mm and 5~mm $\times$ 21~mm, respectively. 
	
	The fact that the coated disks have been obtained by carving a single block of material
	implies that the bonding between disk and coating is perfect and that the sample is obtained without introducing any residual stress, so that detachments 
	are excluded until failure. Note that failure cannot be investigated with the 
	proposed experimental set-up, because it involves out-of-plane buckling and
	may break the polariscope. 

For comparison, both the interior disk and the external coating of the coating/disk complex have also been machined into separate pieces, shown in Fig. \ref{sample_Geom} on the center and on the right. In this way, the 
	inner disk, the annular beam representing the coating, and the coated disk have all been individually tested.

Vertical load has been quasi-statically increased through compression against two horizontal steel plates, equipped with a loading cell (TH-KN2D load cell RC 20 kN, from Gefran) to 
measure (signal was acquired with a NI CRio interfaced via software Labview, ver. 2018 from National Instruments) the applied load, under displacement control (imposed through an electromechanical testing machine, ELE Tritest 50, by ELE International Ltd). The samples under loading have been analyzed with a linear and circular polariscope 
(with quarterwave retarders for 560nm, dark field arrangement; equipped with a white and sodium vapor lightbox at $\lambda$ = 589.3nm, purchased from Tiedemann \& Betz), designed and manufactured at the Instabilities Lab of the University of Trento. 

Photos at white and monochromatic light were taken with a Nikon D200 digital camera, equipped with a AF-S micro
Nikkor (105 mm, 1:2.8G ED) or with a AF-S micro Nikkor (70180 mm, 1:4.55.6 D) lens. 
Monitored with a thermocouple connected to a Xplorer GLX Pasco$^\copyright$, temperature
near the samples during experiments was found to lie around 22.5 $^\circ$C, without sensible oscillations.

\subsubsection{Analysis of the experimental results}
\begin{figure}[htb!]
			\centering
				\includegraphics[scale=1]{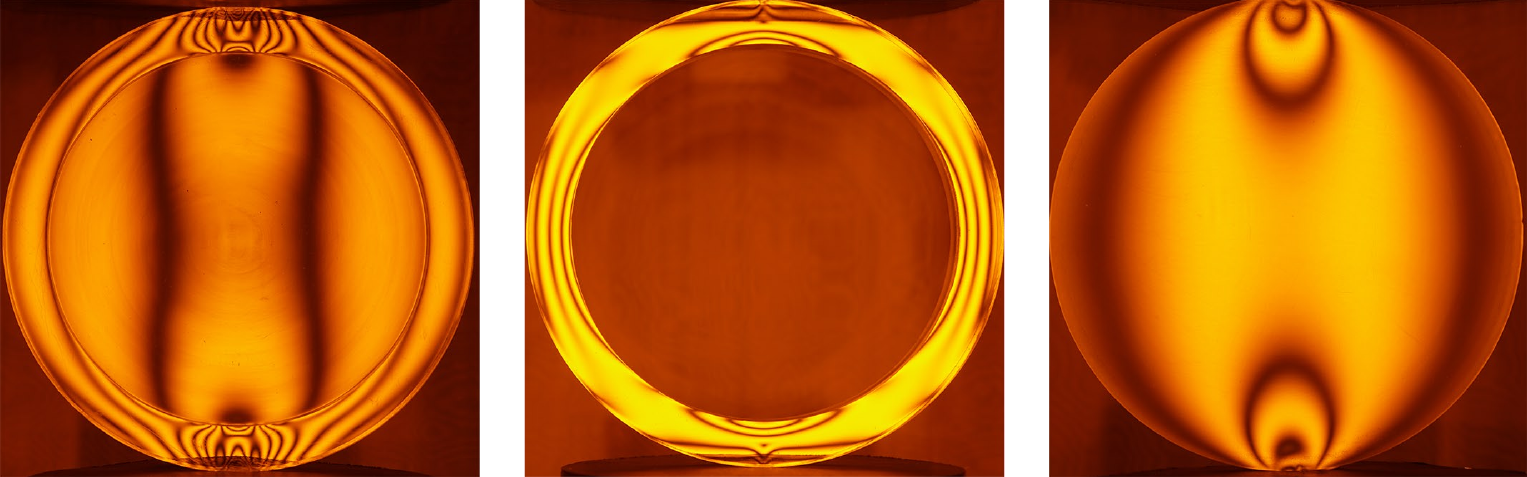} 
			\caption{Monochromatic photoelastic fringes under diametrical compression of: a coated disk (left), its external ring (center), and the inner disk without coating (right). Samples are shown in Fig. \ref{sample_Geom} (upper part).}
			\label{photoelasticityMc}
		\end{figure}

\begin{figure}[htb!]
			\centering
            \includegraphics[scale=1]{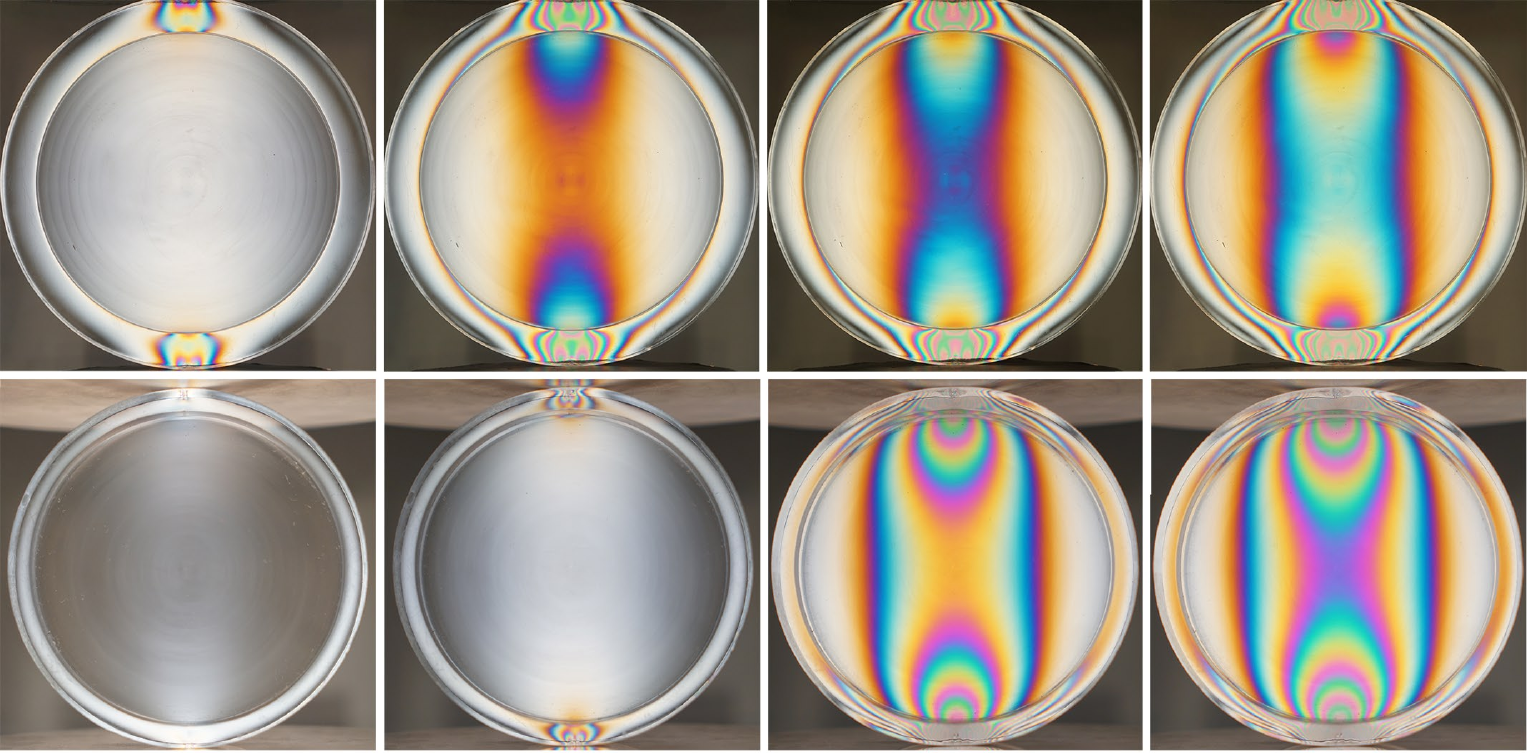} 
			\caption{Photoelastic fringes generated during (vertical) diametrical compression in the coated disks with bending stiffness of the coating EJ$_{1}$=1.4\,$\cdot$10$^{-2}$\,kNm$^{2}$ (upper part) and EJ$_{2}$=7\,$\cdot$10$^{-4}$\,kNm$^{2}$ (lower part). Four increasing values of compression are reported: 3.0 kN (1 kN), 7.0 kN (2.5 kN), 8.5 kN (7.0 kN), and 10.0 kN (8.5 kN) in the upper part (lower part), for the samples shown in Fig. \ref{sample_Geom}; photos have been taken at white circularly polarized light.}
			\label{photoelasticity}
		\end{figure}

Photos taken during the tests at increasing values of loading are reported 
in 
Fig. \ref{photoelasticityMc} at monochromatic light, and  
in Fig. \ref{photoelasticity}
at white light 
(for the coated disk samples reported in Fig. \ref{sample_Geom}). 

Starting the discussion with Fig. \ref{photoelasticityMc} (referred to the samples visible in Fig. \ref{sample_Geom}, upper part), it should be 
noted that the loads for the three samples 
are different, higher for the coated disk (on the left, 10~kN), low for the 
ring (central part, 0.6~kN), and 
intermediate for the uncoated disk (on the right 4.6 kN). This choice has been guided by the fact that at the value used for the coated disk the ring would break and the uncoated disk would suffer an out-of-plane buckling, while at the value used for the ring the coated disk is loaded so little that the fringes are barely visible.  

Experiments on the uncoated disk are 
well-known \cite{frocht1965photoelasticity}, so that only one additional is reported in Appendix \ref{diskApp}, used to calibrate the other photoelastic analyses.  

Further experimental results on the ring, together with comparisons with the calculated stress state, are deferred to Appendix \ref{anelloApp}. 
These photoelastic analyses show that the ring behaves as a circular Euler-Bernoulli beam, where flexural deformation prevails. Therefore, the coating can be modelled with excellent approximation as inextensible, because the ratio between flexural and 
axial stiffnesses becomes negligible.
To substantiate this statement with an evaluation, an annular beam of radius $R$ is assumed to be both flexurally and axially deformable and subject to two diametrical  
forces $F$. The shortening of the diameter of the rod can be written as
\begin{equation}
\Delta = \frac{(\pi^2-8)\,R^3}{4\pi \, EJ}
\left[
1+ 
\underbrace{
\frac{\pi^2}{\pi^2-8} \left( \frac{\rho}{R}\right)^2
}_{axial~deformability}\right],
\end{equation}
where $\rho$ is the radius of inertia of the cross section of the beam. Considering the data pertaining to our experiments, Fig. \ref{sample_Geom}, effects related to the axial deformability of the coating can be evaluated as 0.024 and 0.009, to be compared with 1, representing the flexural deformability.

The fringe pattern reported in  
Fig. \ref{photoelasticityMc} evidences  
the deep effect connected to the presence of the coating, which introduces a nonlocal distribution of the external load, thus strongly affecting the stress state in the disk.  

Considering now Fig. \ref{photoelasticity} 
the photographs refer to four different values 
of loading, namely, 3.0 kN (1 kN), 7.0 kN (2.5 kN), 8.5 kN (7.0 kN), 
and 10.0 kN (8.5 kN) in the upper part (in the lower part).

Although the stress states for the two experiments reported in the figure are qualitatively similar, it is clear that the effect of coating is strongly increased in the case of thick coating (upper part of the figure). 
The fringes visualized in the above reported  photoelastic experiments provide a measure of the difference between the in plane principal stresses (for more details see \cite{frocht1965photoelasticity})
	\begin{equation}
	\label{principalStress}
	\begin{aligned}
		\left|\sigma_{I} - \sigma_{II}\right|={\sqrt{\left(\sigma_{11}-\sigma_{22}\right)^{2}+4\sigma_{12}^{2}}} ,
	\end{aligned}
	\end{equation}
thus permitting a comparison with the analytical solution derived in the previous Sections, that will be applied to a loading distribution modelling the load during the experiments and considered in the following Sections. The comparison, anticipated in Fig. \ref{coatedDisk}, shows that the analytical solution excellently models the experimental results.

	\subsection{An analytical solution for the coated disk modelling the experiments}
	
    In order to apply the general solution for the coated disk obtained in Section \ref{coatdiskSol}, the traction distribution acting on the external portion
	of the coating has to be modelled and implemented. These are the subjects of the present section. 
	
	\subsubsection{Model for the external load applied on the coated disk}\label{lanczosSec}

	The external load will be modelled as a constant normal traction $p$ ($q=0$) applied along two diametrically opposed small arcs of equal amplitude, $s=\alpha \, R$, 
	where $\alpha$ is an angle centred at the vertical diameter of the disk. Using the  complex Fourier series expansion \eqref{p+q*}, multiplying both sides of it by 
		$e^{-mi\theta},\,\,m=\pm{1},\pm{2},...\pm{n}$ and integrating over the whole circle leads to 
			\begin{equation}
	\label{intSigma}
		\int_{\frac{\pi}{2}-\alpha}^{\frac{\pi}{2}+\alpha}p\,e^{-mi\theta}d\theta+\int_{\frac{3\pi}{2}-\alpha}^{\frac{3\pi}{2}+\alpha}p\,e^{-mi\theta}d\theta=\int_{0}^{2\pi}{\left[\sum_{n=1}^{\infty}{D_{-n}e^{-\left(n+m\right)i\theta}}+\sum_{n=0}^{\infty}{D_{n}e^{-\left(n-m\right)i\theta}}\right]d\theta} .
	\end{equation}
	
	The right hand side of equation \eqref{intSigma} is different from zero and equal to $2\pi$ if and only if $n=m$. 
	Hence, at every fixed $m$, one non-null coefficient is determined. Collecting terms with the same power of $e^{\pm{ni\theta}}$ and inverting equation \eqref{intSigma} yield the values of all  coefficients $D_{\pm{n}}$. 
	
	\begin{figure}[hbt!]
		\centering
		\includegraphics[keepaspectratio]{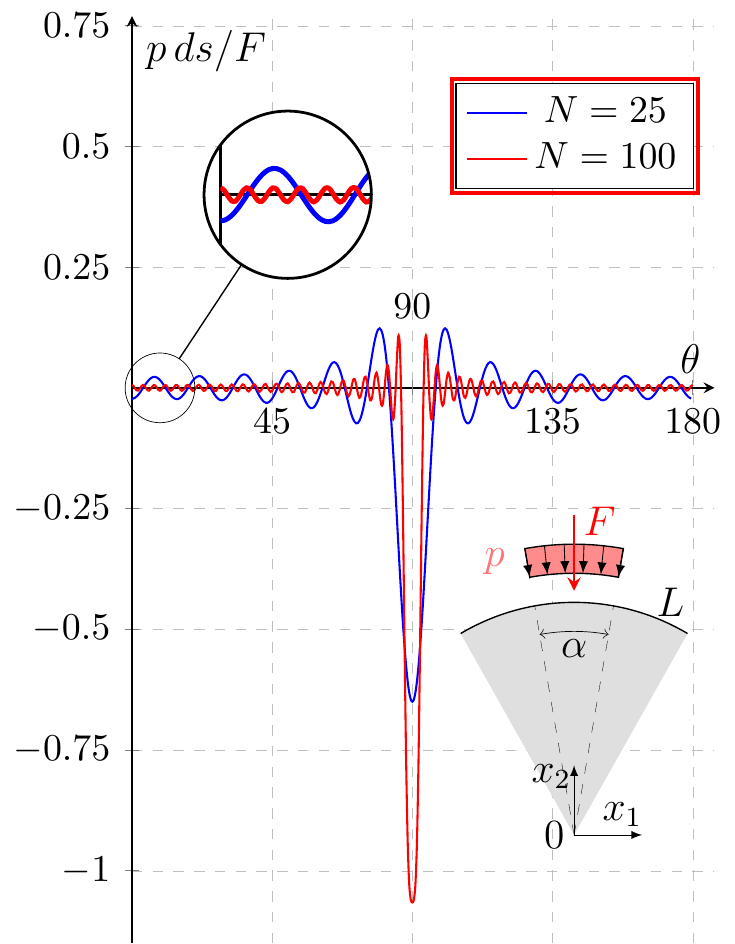} 
		\includegraphics[keepaspectratio]{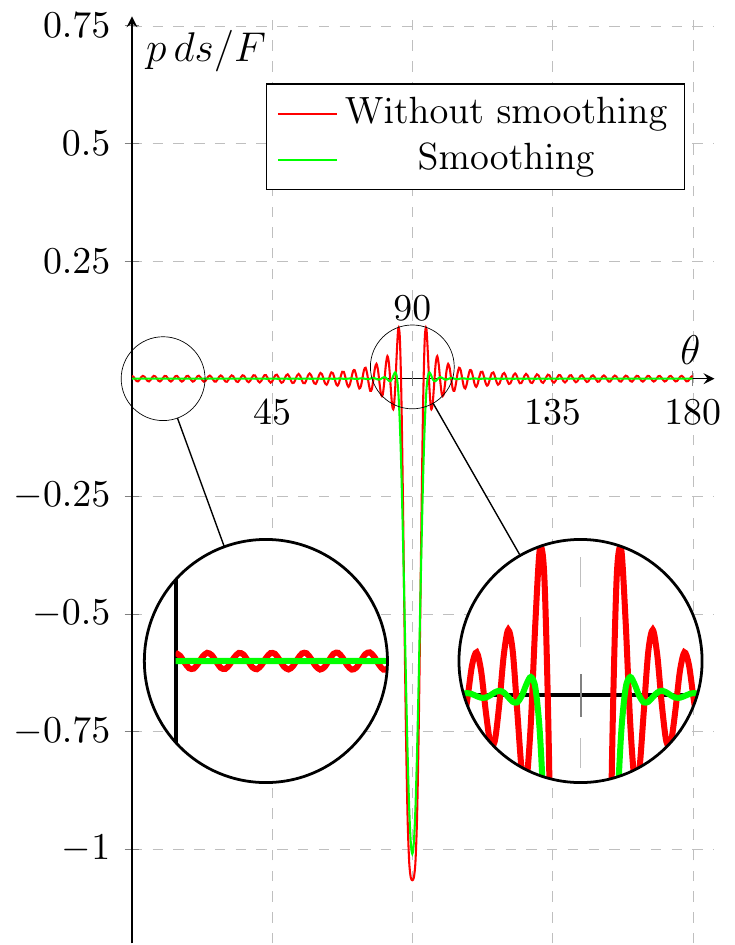}
		\caption{Convergence of the series approximation for the applied load, equation \eqref{p+q*}, using $N=25$ and $N=100$ complex coefficients $D_{\pm{n}}$ (left). Severe oscillations are evident, as related to the presence of a discontinuity in the applied loads, 
		which should vanishes at $\theta=\pi/2\mp{\alpha}$ and $\theta=3\pi/2\pm{\alpha}$. Both global and local oscillation can be filtered out by means of a smoothing technique (right) resulting in a better approximation for the applied external load using $N=100$ complex coefficients $D^{*}_{\pm{n}}$ having the form of equation  \eqref{Dstar}}.
		\label{Series95}
	\end{figure}
	
	The convergence of the series approximation for an applied load of unit resultant is analyzed in Fig. \ref{Series95} (central part). 
	A satisfactory representation for the applied concentrated load can be obtained with an arc length $\alpha=5\degree$ and a truncation of the complex Fourier series at $N=100$. The convergence of the series is analyzed in Fig. \ref{Series95}  
	(on the left), where results are reported for  $N=25$ and $N=100$. Here it may be concluded that: (i.) the maximum load does not exceed the unit value up to $N=100$ (for $N=25$ the maximum value is $\approx 0.65$) and (ii.) severe sign variations of the represented function are visible, a phenomenon known as \lq Gibbs oscillations'.
	
	Although the approximation obtained with $N=100$ is judged as satisfactory and will be used in the following, the local smoothing technique proposed by Lanczos (1966) has 
	been also implemented to further increase the precision in the representation. 
The Lanczos technique can be explained with reference to a given a Fourier series representation of a function $f\left(\theta\right)$, which displays a jump discontinuity, with partial sum 
	\begin{equation}
	\label{peretta}
	f_{N}\left(\theta\right)=a_{0}+\sum_{n=1}^{N}{a_{n}\cos{\left(n\theta\right)}}+\sum_{n=1}^{N}{b_{n}\sin{\left(n\theta\right)}} .
	\end{equation}
	Lanczos (1966) introduced the smoothed value for the partial sum (\ref{peretta}) as 
	\begin{equation}
	\label{LanczosSum}
		f^{*}_{N}\left(\theta\right)=a^{*}_{0}+\sum_{n=1}^{N}{\frac{\sin{\left(n\pi/N\right)}}{n\pi/N}\left[a^{*}_{n}\cos{\left(n\theta\right)}+b^{*}_{n}\sin{\left(n\theta\right)}\right]} .
	\end{equation}
Using now the complex notation, 
	\begin{equation}
	\begin{aligned}
		\cos{n\theta}=\frac{1}{2}\left(e^{i\,n\theta}+e^{-i\,n\theta}\right) ,&&
		\sin{n\theta}=\frac{1}{2i}\left(e^{i\,n\theta}-e^{-i\,n\theta}\right) ,
	\end{aligned}
	\end{equation}
	the partial sum expressed by equation \eqref{LanczosSum} can be rewritten as 
	\begin{equation}
		f^{*}_{N}\left(\theta\right)=a^{*}_{0}+\sum_{n=1}^{N}{\frac{\sin{\left(n\pi/N\right)}}{n\pi/N}\left[\frac{a^{*}_{n}}{2}\left(e^{i\,n\theta}+e^{-i\,n\theta}\right)+\frac{b^{*}_{n}}{2i}\left(e^{i\,n\theta}-e^{-i\,n\theta}\right)\right]} ,
	\end{equation}
	so that by collecting terms yields 
	\begin{equation}
	\label{LanczosCmplx}
		f^{*}_{N}\left(\theta\right)=a^{*}_{0}+\sum_{n=1}^{N}{\frac{\sin{\left(n\pi/N\right)}}{n\pi/N}\left[\left(\frac{a^{*}_{n}-i\,b_{n}^{*}}{2}\right)e^{in\theta}+\left(\frac{a^{*}_{n}+i\,b_{n}^{*}}{2}\right)e^{-in\theta}\right]} ,
	\end{equation}
	and adopting the notation used in equation \eqref{gtau} finally leads to 
	\begin{equation}
	\label{fStarN}
		f^{*}_{N}\left(\theta\right)=a^{*}_{0}+\sum_{n=1}^{N}{\frac{\sin{\left(n\pi/N\right)}}{n\pi/N}\left[\left(\frac{a^{*}_{n}-i\,b_{n}^{*}}{2}\right)g^{-n}\left(\tau\right)+\left(\frac{a^{*}_{n}+i\,b_{n}^{*}}{2}\right)g^{n}\left(\tau\right)\right]}.
	\end{equation}
	
	Equations \eqref{fStarN} and \eqref{intSigma} allow the determination of the new coefficients $D_{n}^{*}$ for the smoothed function 
	\begin{equation}
	\label{Dstar}
	\left\{\begin{array}{l}
		D_{0}^{*}=a_{0}^{*} ,\\[10pt]
		\displaystyle D_{-n}^{*}=\left(\frac{a_{n}+ib_{n}}{2}\right)\frac{\sin{\left(n\pi/N\right)}}{n\pi/N} ,\\[10pt]
		\displaystyle  D_{n}^{*}=\left(\frac{a_{n}-ib_{n}}{2}\right)\frac{\sin{\left(n\pi/N\right)}}{n\pi/N} ,
	\end{array}\right.
	\end{equation}
	providing a sufficiently accurate representation of the applied load, for appropriate choice of $N$, as shown in Fig. \ref{Series95} on the right.

	\subsubsection{Solution for the coated disk subject to opposite force distributions \label{coatSec}}
	
	All ingredients are now ready to obtain the solution of a coated disk subjected to two equal and opposite distributions of surface tractions of the type shown in Fig. \ref{Series95}. 
	This solution is derived below and will be used for  comparison with the experimental results. 
	
	A substitution of equations \eqref{Ak=A-kconj} and \eqref{A-k+1(Df)} into equation \eqref{potentialdisczc}, provides the complex potentials for the coated disk when subjected to two opposite radial traction distributions, written in terms of the coefficients $D_{\pm{n}}$
	\begin{equation} 
	\label{potentialdiskfD}
	\begin{aligned}
		&\varphi\left(z\right)=-\mu{R^{4}}\sum_{n=2}^{\infty}{\left\{\frac{\left(n-1\right)\overline{D_{-n}}+\left(n+1\right)D_{n}}{\left(n+1\right)\left[\kappa{EJ}n^{2}\left(n^{2}-1\right)+R^{3}\mu{b}\left(n+\kappa{n}+\kappa-1\right)\right]}\right\}}\,g^{-\left(n+1\right)}\left(z\right) ,\\[5pt] 
		&\psi\left(z\right)=\mu{R^{4}}\sum_{n=2}^{\infty}{\left\{\frac{\left(n-1+\kappa\right)\left[\left(n-1\right)\overline{D_{-n}}+\left(n+1\right)D_{n}\right]}{\left(n-1\right)\left[\kappa{EJ}n^{2}\left(n^{2}-1\right)+R^{3}\mu{b}\left(n+\kappa{n}+\kappa-1\right)\right]}\right\}}\,g^{-\left(n-1\right)}\left(z\right) ,\\ 
	\end{aligned}
	\end{equation}
	so that the displacement follows as 
	\begin{equation} 
	\label{u_coated}
	\begin{aligned}
		{u}\left(z\right)=&-\frac{R^{4}\kappa}{2}\sum_{n=2}^{\infty}\left\{\frac{\left[\left(n-1\right)\overline{D_{-n}}+\left(n+1\right)D_{n}\right]}{\left(n+1\right)\left[\kappa{EJ}n^{2}\left(n^{2}-1\right)+R^{3}\mu{b}\left(n+\kappa{n}+\kappa-1\right)\right]}\,g^{-\left(n+1\right)}\left(z\right)\right.\\
		&\left.-\frac{\left[\left(n-1\right){D_{-n}+\left(n+1\right)\overline{D_{n}}}\right]\left[r^{2}\left(n-1\right)-R^{2}\left(n-1+\kappa\right)\right]R^{-2n}}{\kappa\left(n-1\right)\left[\kappa{EJ}n^{2}\left(n^{2}-1\right)+R^{3}\mu{b}\left(n+\kappa{n}+\kappa-1\right)\right]r^{-2\left(n-1\right)}}\,g^{n-1}\left(z\right)\right\} ,
	\end{aligned}
	\end{equation}
	and the stress components, equation \eqref{muskhelishvilifields}$_{2}$,  as
	\begin{equation} 
	\label{sA_coated}
	\centering
	\begin{aligned}
		\frac{\sigma_{11}+\sigma_{22}}{4\mu{R^{3}}}=- \sum_{n=2}^{\infty}{\mathrm{Re}\!\left(\frac{\left[\left(n-1\right)\overline{D_{-n}}+\left(n+1\right)D_{n}\right]}{\kappa{EJ}n^{2}\left(n^{2}-1\right)+R^{3}\mu{b}\left(n+\kappa{n}+\kappa-1\right)}\,g^{-n}\left(z\right)\right)} , 
	\end{aligned}
	\end{equation}
	while, from equation \eqref{muskhelishvilifields}$_{3}$, as
	\begin{equation} 
	\label{sB_coated}
	\begin{aligned}
		\frac{\sigma_{22}-\sigma_{11}+2i\sigma_{12}}{2\mu{R}}= &\sum_{n=2}^{\infty}{\frac{
		\left[\left(n-1\right)\overline{D_{-n}}+\left(n+1\right)D_{n}\right]\left[R^{2}\left(n+\kappa-1\right)-r^{2}n\right]}{\kappa{EJ}n^{2}\left(n^{2}-1\right)+R^{3}\mu{b}\left(n+\kappa{n}+\kappa-1\right)}\,g^{-\left(n-2\right)}\left(z\right)} .
	\end{aligned}
	\end{equation}
	
	The in-plane deviatoric stress can be calculated using equation \eqref{principalStress} from the knowledge of 
	$\sigma_{11}$, $\sigma_{22}$, and $\sigma_{12}$.

The results are reported in Fig. \ref{coatedDisk},  where they are compared with the photoelastic experiments performed on the coated disks, shown in Fig. \ref{sample_Geom}.
	The internal forces in the coating, equations \eqref{azione}, are completely determined by the displacement component $u_{r}$ in terms of the coefficients $A_{1-n}$ in equation  \eqref{A-k+1(Df)}. Recalling equations  \eqref{A2} and \eqref{Ak=A-kconj}, the derivatives of complex displacement involved in equation \eqref{azione} are found as
	\begin{equation}
	\label{beamud}
	\begin{aligned}
	&\frac{du_{r}}{ds}=-\frac{1}{R}\sum_{n=1}^{\infty}{\mathrm{Im}\!\left(-\left(n+1\right)A_{-n}\,g^{n+1}\left(\tau\right)+\left(n-1\right)A_{n}\,g^{-\left(n-1\right)}\left(\tau\right)\right)} ,\\ 
	&\frac{d^{2}{u_{r}}}{ds^{2}}=-\frac{1}{R^{2}}\sum_{n=1}^{\infty}{\mathrm{Re}\!\left(\left(n+1\right)^{2}A_{-n}\,g^{n+1}\left(\tau\right)+\left(n-1\right)^{2}A_{n}\,g^{-\left(n-1\right)}\left(\tau\right)\right)} ,\\ 
	&\frac{d^{3}u_{r}}{ds^{3}}=\frac{1}{R^{3}}\sum_{n=1}^{\infty}{\mathrm{Im}\!\left(-\left(n+1\right)^{3}A_{-n}\,g^{n+1}\left(\tau\right)+\left(n-1\right)^{3}A_{n}\,g^{-\left(n-1\right)}\left(\tau\right)\right)} ,\\ 
	&\frac{d^{4}u_{r}}{ds^{4}}=\frac{1}{R^{4}}\sum_{n=1}^{\infty}{\mathrm{Re}\!\left(\left(n+1\right)^{4}A_{-n}\,g^{n+1}\left(\tau\right)+\left(n-1\right)^{4}A_{n}\,g^{-\left(n-1\right)}\left(\tau\right)\right)} . 
	\end{aligned}
	\end{equation}
	
	A substitution of equations \eqref{beamud} into expressions \eqref{azione} leads to the internal forces along the coating as  
	\begin{equation}
	\label{beamAzione}
	\begin{aligned}
	&M=-\frac{EJ}{R^{2}}\sum_{n=1}^{\infty}{n\,\mathrm{Re}\!\left(\left(n+2\right)A_{-n}\,g^{n+1}\left(\tau\right)+\left(n-2\right)A_{n}\,g^{-\left(n-1\right)}\left(\tau\right)\right)}  ,\\ 
	&T=-\frac{EJ}{R^{3}}\sum_{n=1}^{\infty}{n\,\mathrm{Im}\!\left(-\left(n+1\right)\left(n+2\right)A_{-n}\,g^{n+1}\left(\tau\right)+\left(n-1\right)\left(n-2\right)A_{n}\,g^{-\left(n-1\right)}\left(\tau\right)\right)}  ,\\ 
	&N=-\frac{EJ}{R^{3}}\sum_{n=1}^{\infty}{n\,\mathrm{Re}\!\left(\left(n+2\right)\left(n+1\right)^{2}A_{-n}\,g^{n+1}\left(\tau\right)+\left(n-2\right)\left(n-1\right)^{2}A_{n}\,g^{-\left(n-1\right)}\left(\tau\right)\right)}\\
	&\,-R\,\mathrm{Re}\!\left(\sum_{n=1}^{\infty}{\left[D_{-n}+\frac{2\mu{b}}{\kappa{R}}\left(n-1\right)A_{1-n}\right]g^{n}\left(\tau\right)+\sum_{n=0}^{\infty}{\left[D_{n}+\frac{2\mu{b}}{R}\left(n-1\right)\overline{A_{1-n}}\right]g^{-n}\left(\tau\right)}}\right) . 
	\end{aligned}
	\end{equation}
	
\begin{figure}
		\centering
		\subfigure{
		\includegraphics[keepaspectratio]{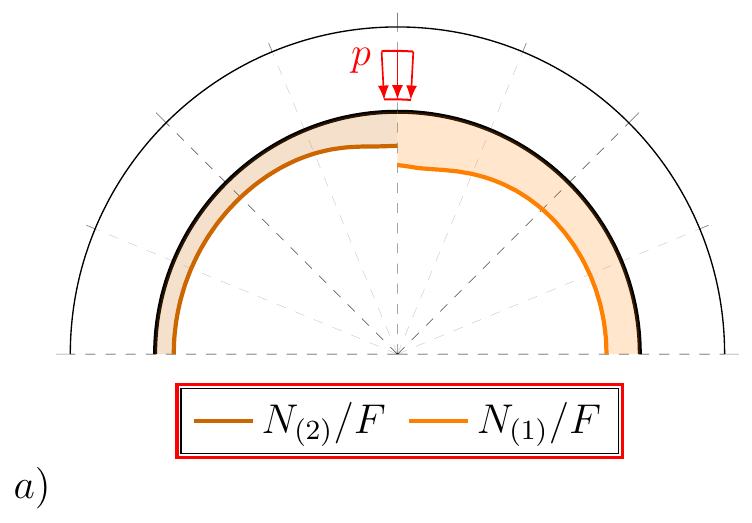} 
		\includegraphics[keepaspectratio]{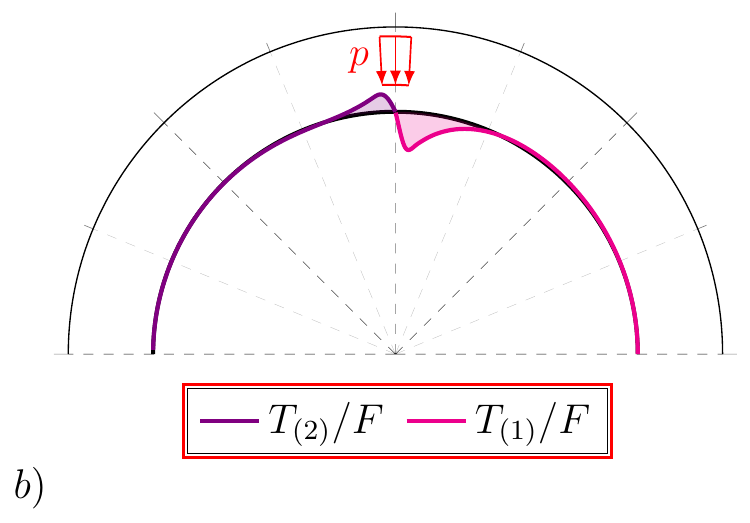} 
		}
		\subfigure{
		\includegraphics[keepaspectratio]{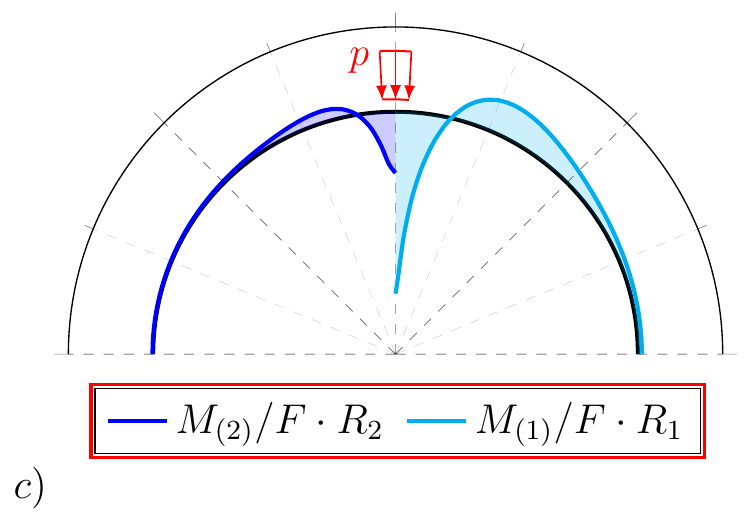} 
		\includegraphics[keepaspectratio]{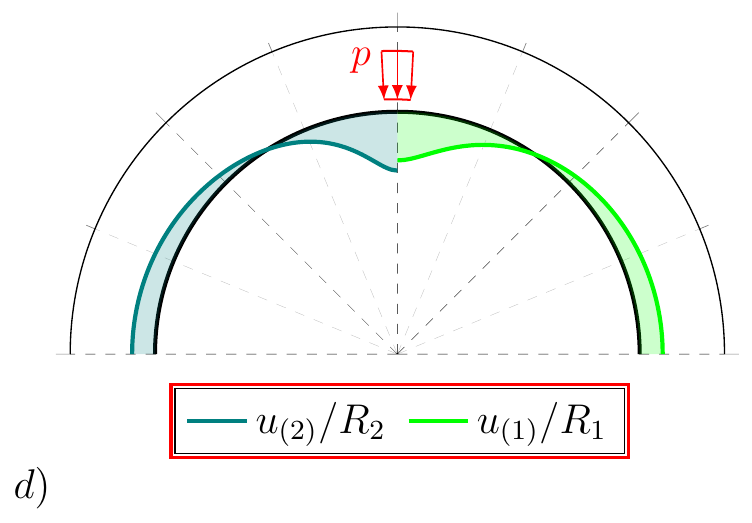} 
		}
		\caption{
		Force distributions and displacements of the coating for the two coated disks used in the experiments and loaded by two equal and opposite force  distributions (reported in red, with resultant $F=7$kN). The coatings are characterized by 
		bending stiffness  EJ$_{1}$=1.4\,$\cdot$10$^{-2}$\,kNm$^{2}$ (index 1) and EJ$_{2}$=7\,$\cdot$10$^{-4}$\,kNm$^{2}$ (index 2), representative of the samples reported on the left of Fig. \ref{sample_Geom}. 
		(a) Normalized normal forces $N_{(1)}/F$, $N_{(2)}/F$; (b) shear forces $T_{(1)}/F$, $T_{(2)}/F$; (c) normalized bending moments $M_{(1)}/(F\cdot{R_{1}})$, $M_{(2)}/(F\cdot{R_{2}})$, equations \eqref{beamAzione};  (d) 
		Normalized displacements $u_{(1)}/R_{1}$ and $u_{(2)}/R_{2}$, equation \eqref{u_coated} along the external surface of the coated disk. }
		\label{coat}
	\end{figure}

The two coatings of the disks 
used in the experiments are characterized by two strongly different bending stiffnesses, namely, EJ$_{1}$=1.4\,$\cdot$10$^{-2}$\,kNm$^{2}$ and EJ$_{2}$=7\,$\cdot$10$^{-4}$\,kNm$^{2}$. 
Analytical results referred to these two rings, coating a disk, are reported in Fig. \ref{coat} (only a quarter of the coating is shown) 
and discriminated with the 
indices 1 and 2, referring 
to EJ$_1$ and EJ$_2$, respectively. Normal, $N$, and 
shear, $T$, forces and bending moments, $M$  are reported 
in panels (a), (b), and (c), as calculated from equations \eqref{beamAzione}. 
The displacement of the 
coating $u$, equation \eqref{u_coated}, is reported in Panel (d). 

Fig. \ref{coat} shows that the internal forces and the displacement are quantitatively, but not qualitatively, affected by the stiffness of the coating. The displacement of the coating evidences that a 
stiff coating leads to a higher non-locality effect. 

 From the knowledge of the internal forces, $N$, $T$, and $M$, the state of stress at every point of the cross section of the coating can be obtained by considering the model of an annular beam \cite{timoshenko1950strength}. From the stress distribution, the in-plane deviatoric stress $\sigma_{I}-\sigma_{II}$ in the coating can be obtained, as reported inside the coatings visible in Figs. \ref{coatedDisk} and \ref{diskDiff}. 
 
 Additional deviatoric stress distributions, determined inside the annular beams so far considered, but taken as isolated, are reported in Appendix \ref{anelloApp}, where they are compared to the results from photoelastic experiments.

\newpage

	\subsubsection{Special case: the disk with a coating imposing only an isoperimetric constraint, but not transmitting bending}
	
		At vanishing $EJ$ for the coating, equation \eqref{A-k+1(Df)} reduces to the solution of a disk coated with a device (which may be imagined as an inextensible string) maintaining the isoperimetric constraint.   
		
	The determination of the elastic fields inside a disk coated with the above-mentioned axially inextensible constraint 
	can be pursued by using equations \eqref{muskhelishvilifields} for the coated disk, in the limit for $EJ\to{0}$. 
	In particular, the Kolosov-Muskhelisvili potentials $\varphi$ and $\psi$, equations \eqref{potentialVarphi}$_{1}$ and \eqref{potentialPsi}$_{1}$, assume the same expressions \eqref{potentialdiskfD} as for the coated case, but now the new coefficients $A_{1-n}^{disk}$ for the disk without coating will be generated by requiring that the bending stiffness $EJ$ in equation \eqref{A-k+1(Df)} vanishes
	\begin{equation} 
	\label{A-k+1Disk}
	\begin{aligned}
		A_{1-n}^{disk}=\lim_{EJ\to{0}}{A_{1-n}}=-\frac{\kappa{R}}{2\mu{b}}\frac{\left(n+1\right)\overline{D_{n}}+\left(n-1\right)D_{-n}}{\left(n-1\right)\left(n+\kappa{n}+\kappa-1\right)} ,\,&&\text{for}\,n\ge{2} ,
	\end{aligned}
	\end{equation}
	and hence potentials, equations \eqref{potentialdiskfD}, assume the form 
	\begin{equation} 
	\label{potentialNoCoat}
	\begin{aligned}
		&\varphi^{disk}\left(z\right)=-R\sum_{n=2}^{\infty}{\left\{\frac{\left(n-1\right)\overline{D_{-n}}+\left(n+1\right)D_{n}}{b\left(n+1\right)\left(n+\kappa{n}+\kappa-1\right)}\right\}}\,g^{-\left(n+1\right)}\left(z\right) ,\\[5pt] 
		&\psi^{disk}\left(z\right)=R\sum_{n=2}^{\infty}{\left\{\frac{\left(n-1+\kappa\right)\left[\left(n-1\right)\overline{D_{-n}}+\left(n+1\right)D_{n}\right]}{b\left(n-1\right)\left[\left(n+\kappa{n}+\kappa-1\right)\right]}\right\}}\,g^{-\left(n-1\right)}\left(z\right) .\\ 
	\end{aligned}
	\end{equation}
	
	All the elastic fields which solve the problem of the disk with the isoperimetric  coating without bending stiffness can now be computed through relations \eqref{muskhelishvilifields}, with the substitutions $\varphi\rightarrow\varphi^{disk},\psi\rightarrow\psi^{disk}$.
	
	\subsubsection{Stress distributions for different models of coating}
	
\begin{figure}[hbt!]
			\centering
				\includegraphics[keepaspectratio, scale=1]{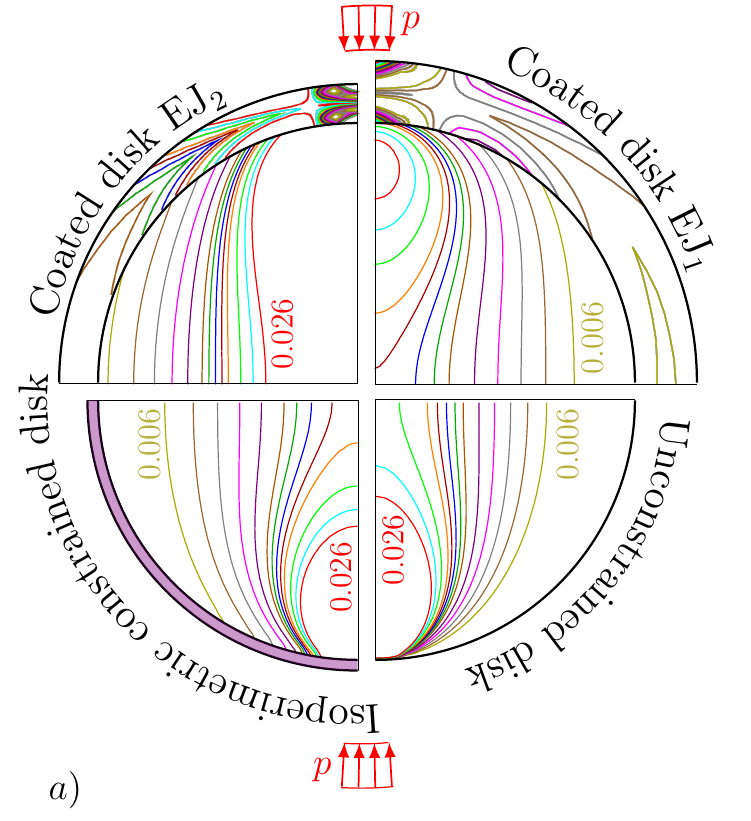}
				\hspace{2.9 pt}
				\includegraphics[keepaspectratio, scale=1]{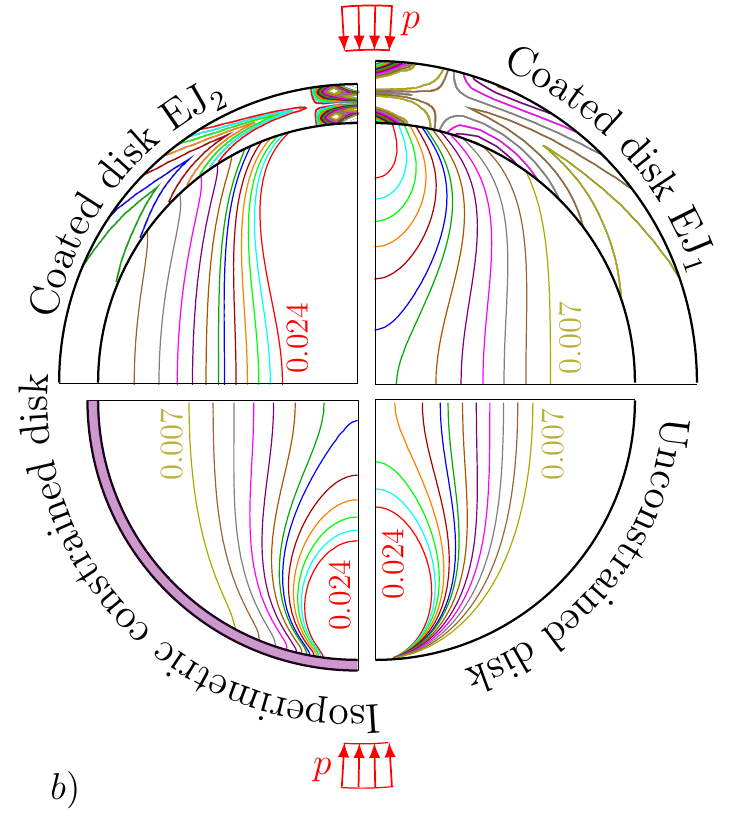}
			\caption{
			Various models of disks loaded under diametrical compression: with two coatings stiff under bending,  EJ$_{1}$, 
			and 
		EJ$_{2}$; with a coating imposing an isoperimetric constraint, but without bending stiffness; without any coating. 
		Dimensionless stress contours for (a) in-plane deviatoric stress and (b) normalized von Mises stress are reported for disk and coating 
		(sketched as a purple line when it corresponds to a mere isoperimetrical constraint). 
		The coating introduces a non local stress diffusion, an  effect which tends to vanish when the bending stiffness of the coating is decreased, is strongly reduced for a disk with isoperimetric constraint, and vanishes when boundary constraints on the disk are absent. }
			\label{diskDiff}
		\end{figure}

	Contour plots for 
	the in-plane principal deviatoric stress  $\left|\sigma_{I}-\sigma_{II}\right|$ and for the von Mises stress 
    are reported inside the disk in Fig. \ref{diskDiff}, (a) and (b), respectively, both made dimensionless through division by the elastic shear modulus $\mu$. 
	Four models of disks are considered (only a quarter of the domain is reported), all loaded 
	with the two radial compressive stress distributions shown in the figure. 
	Results refer to: two coated disks, with (i.) EJ$_{1}$=1.4\,$\cdot$10$^{-2}$\,kNm$^{2}$ and (ii.) EJ$_{2}$=7\,$\cdot$10$^{-4}$\,kNm$^{2}$ (upper parts); (iii.) the disk coated with the isoperimetric constrain without bending stiffness; (iv.) the \lq nude' disk, namely, unconstrained. 

Note that the deviatoric and von Mises stress distributions are similar, as it may be expected as both are providing measures of distortional stress. Numbers reported on the contours facilitate comparisons and same colors refer to the same stress level. 
	
The figure highlights the role of the bending stiffness of the coating, showing that 
	the stress strongly increases at the mid point of load distribution, when the bending stiffness vanishes, as in the  case of the isoperimetric constrained disk. This effect  is even more pronounced in the case of the unconstrained disk. 
	A non-local diffusion of the stresses are clearly visible in the coated cases.

\subsection{Comparison between photoelastic experiments and mechanical modelling} \label{PhEvSec}

\subsubsection{Determination of stresses from fringe colours}

\begin{figure}[ht!]
	\centering
	\includegraphics[keepaspectratio,scale=1.0]{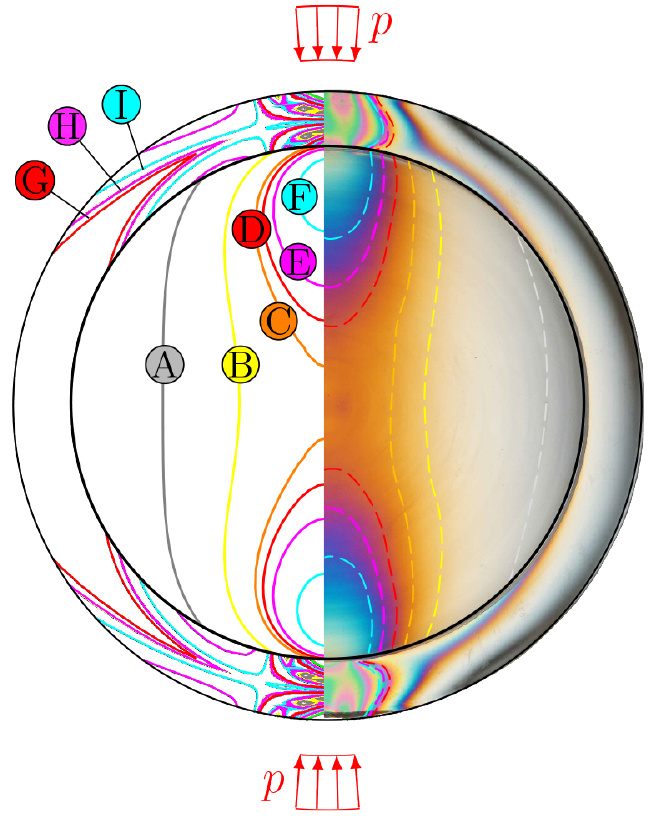} 
	\includegraphics[keepaspectratio, scale=1.0]{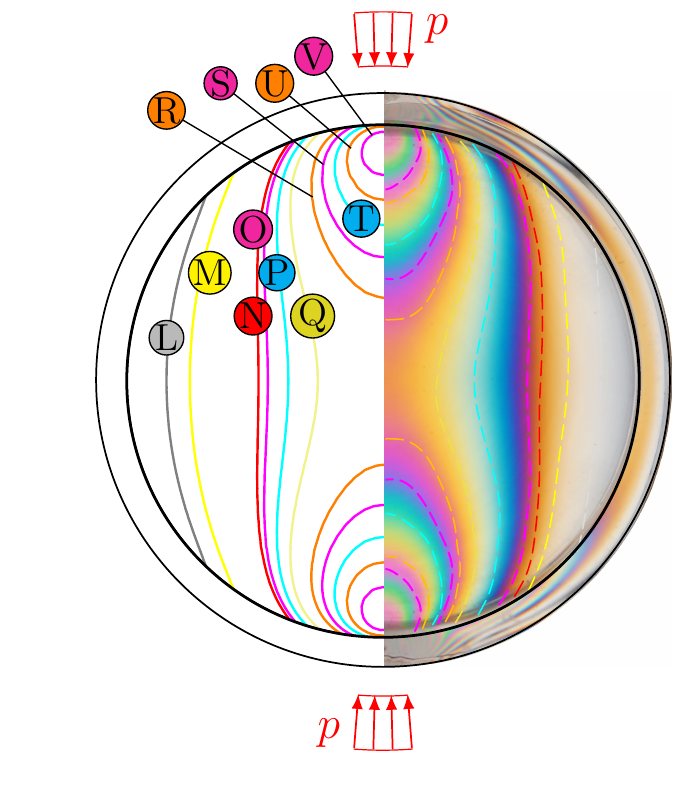} 
	\caption{Determination of the in-plane deviatoric stress from the photoelastic fringes. The dashed lines on the right half of the sample refer to Table \ref{Results2022}, while the location of the same stress level is reported on the left half of the sample, as evaluated with the analytic solution. 
	The two samples are those reported in Fig. \ref{sample_Geom}.
	}
	\label{CoatedFring}
\end{figure}

The photoelastic analysis of the coated disk is reported below, referred to the same vale of the load resultant $F=7$kN, applied at the top of the vertical diameter of the samples shown in Fig. \ref{sample_Geom}. 
Results are reported in Fig. \ref{CoatedFring}, where the photoelastic image is reported on the right half of the sample, while the analytical solution is depicted on the left half. 
The in-plane deviatoric stress as determined from the analysis of the transmitted colours is reported in Table \ref{Results2022}.
\begin{table}[ht!]
\begin{minipage}{\linewidth}
	\centering
		\begin{tabular}{c@{\hspace{3\tabcolsep}}c@{\hspace{2\tabcolsep}}c@{\hspace{2\tabcolsep}}c@{\hspace{1\tabcolsep}}c@{\hspace{2\tabcolsep}}c@{\hspace{2\tabcolsep}}}
			\toprule[1.5pt]
			&       & \multicolumn{2}{c}{\vtop{\hbox{\strut \bf Coated disk sample}\hbox{\strut \small \,\,\,upper part of Fig. \ref{sample_Geom} }}}  & \multicolumn{2}{c}{\vtop{\hbox{\strut \bf Coated disk sample}\hbox{\strut \small \,\,\,lower part of Fig. \ref{sample_Geom} }}} \\
			\cmidrule(rr){3-4}\cmidrule(rr){5-6}
			{\bf{Transmitted color}} & {$\bf N\!\cdot\!\lambda$} & {\bf Line}  & {$\bf\left|\sigma_{I}-\sigma_{II}\right|/\mu$}  & {\bf Line}  & {$\bf\left|\sigma_{I}-\sigma_{II}\right|/\mu$} \\
			{$[-]$} & {$[\text{nm}]$} & {$[-]$} & {$[-]$} & {$[-]$} & {$[-]$}\\ \midrule
			Gray & 1$\cdot$218 & A  & 0.0083  & L & 0.0083 \\
			Brilliant yellow & 1$\cdot$390 & B  & 0.0148 & M  & 0.0148 \\
			Orange & 1$\cdot$505 & C  & 0.0192 & -  & - \\
			Red & 1$\cdot$536 & D  & 0.0203 & N  & 0.0203 \\
			Indigo-violet & 1$\cdot$575 & E  & 0.0219 & O  & 0.0219 \\
			Sky blue & 1$\cdot$664 & F  & 0.0252 & P  & 0.0252 \\
			Brilliant Yellow & 2$\cdot$390 & -  & - & Q  & 0.0296 \\
			Orange & 2$\cdot$505 & -  & - & R  & 0.0384 \\
			Red & 2$\cdot$536 & G  & 0.0108 & -  & - \\
			Indigo-violet & 2$\cdot$575 & H  & 0.0116 & S  & 0.0438 \\
			Sky blue & 2$\cdot$664 & I  & 0.0133 & T  & 0.0505 \\
			Orange & 3$\cdot$505 & -  & - & U  & 0.0576 \\
			Indigo-violet & 3$\cdot$575 & -  & - & V  & 0.0657 \\
			\bottomrule
		\end{tabular}\captionof{table}{\label{Results2022}
		Determination of the in-plane deviatoric stress $\left|\sigma_{I}-\sigma_{II}\right|/\mu$ from the photoelastic experiments performed in the sample reported in the upper part of Fig. \ref{sample_Geom}.
		}
	\end{minipage}
\end{table}

The stress field inside the disk shows an excellent agreement between theory and experiments, while the in-plane deviatoric stress in the external coating 
evidences some discrepancies. The latter are due to the fact that the coating has been modelled as an interface, so that the evaluation of the stresses is conducted starting from the knowledge of the internal force resultants, equations \eqref{beamAzione}, using the classical de Saint-Venant theory, which is an approximation. Moreover, the stress level in the external part of the coating near the zone where the load is applied makes the reading of the transmitted colors difficult. 
For this reason, a quantitative evaluation of the stresses inside the external coating is not reported for the sample in the lower part of Fig. \ref{sample_Geom}.

\begin{figure}[ht!]
		\centering
			\includegraphics[keepaspectratio,scale=1.0]{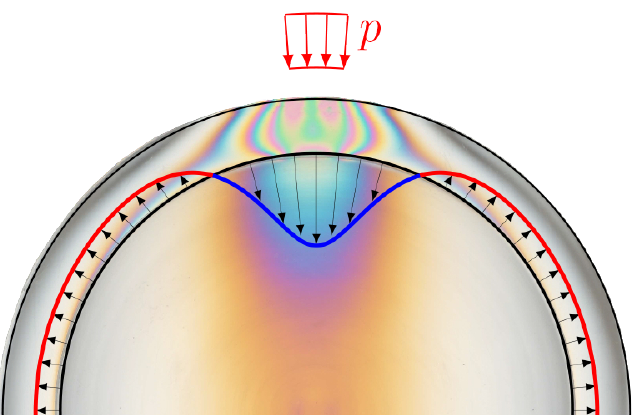}
			\hspace{12pt}
			\includegraphics[keepaspectratio, scale=1.0]{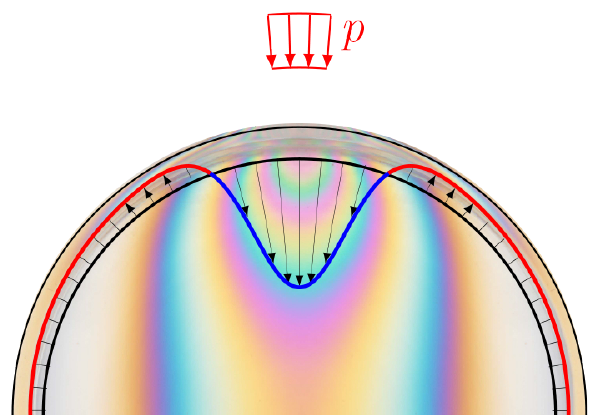} 
		\caption{
		The traction applied to the disk is reported along the interface between coating and disk. When tensile (red line), the traction determines the zone where detachment of coating might occur (it does not in our experiments). The samples are
		those shown in Fig. \ref{sample_Geom}.
		}
		\label{detach}
	\end{figure}

Although detachment is not possible in our experiments, points at the interface where detachment between disk and coating may occur are those subject  to tensile tractions. The distribution of the latter along the disk/coating interface are depicted in Fig. \ref{detach}, where the two samples shown in Fig. \ref{sample_Geom} are subject to the  
	 resultant force $F=7$kN. Note that red line refers to tensile  tractions, while blue refers to compressive.		

\section{Conclusions}

Bulk materials having their external surface coated with a film of  mechanical properties different from the base material are commonly used in a number of technologies. 
Within this context, a paradigmatic mechanical problem has been formulated in the present article, solved, and validated against experiments. This is a linear elastic (isotropic) circular disk, coated
with a thin and stiff film, modelled as a linear elastic circular beam unshearable and axially inextensible, but linearly deformable under bending. The coating therefore imposes an isoperimetric constraint to the
disk and a nonlocal load diffusion on its boundary. The latter effect is related to the introduction of an internal length scale and implies that concentrated forces applied to the external of the coating/disk complex do not introduce a singularity. 
It has been shown that the considered problem can be analytically solved via complex potentials formalism, so that the full displacement, strain, and stress fields have been derived for a generic applied external load. Finally, it has been shown how to practically realize the coated disk with a photoelastic material, thus producing two models which have been used to validate the analytical solution, in the case when two equal and opposed load distributions are applied on a small circular segment. The presented results open new possibilities in the design of coated solids of cylindrical geometry, which may find applications in micro and nano technologies, for instance in the characterization of nanowires via nanoindentation.

\section*{Acknowledgements}

D.B. remembers with great pleasure and emotion the longstanding, intense, fruitful, and beautiful cooperation with Natasha and Sasha on so many topics of science. He looks forward to continue enjoying scientific collaboration with them and sincere friendship for many more years to come. 

D.B. and M.G. gratefully acknowledges the funding from the European Union's Horizon 2020 research and innovation programme under the Marie Sklodowska-Curie grant agreement No 955944-REFRACTURE2.
S.G.M. gratefully acknowledges support from National Science Foundation, award NSF CMMI - 2112894.
\begin{appendices}
	\section{Calibration of the photoelastic model} \label{diskApp}
\begin{figure}[ht!]
	\centering
	\includegraphics[keepaspectratio,scale=1]{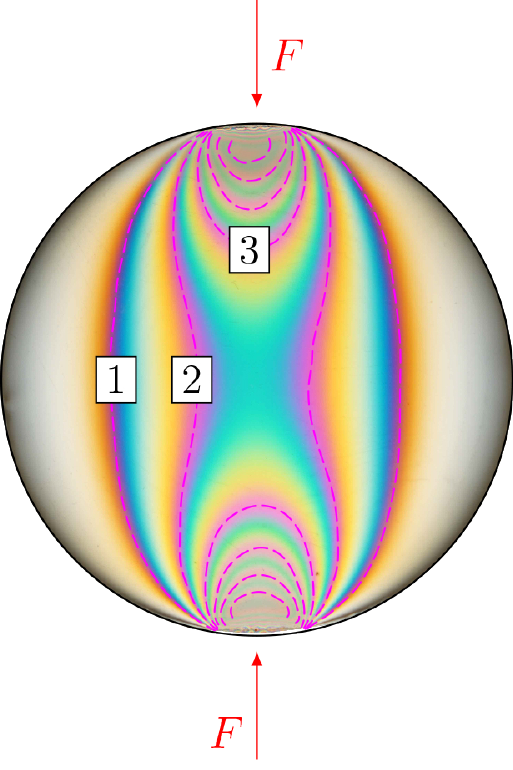}
	\hspace{30.5 pt}
	\includegraphics[keepaspectratio,scale=1]{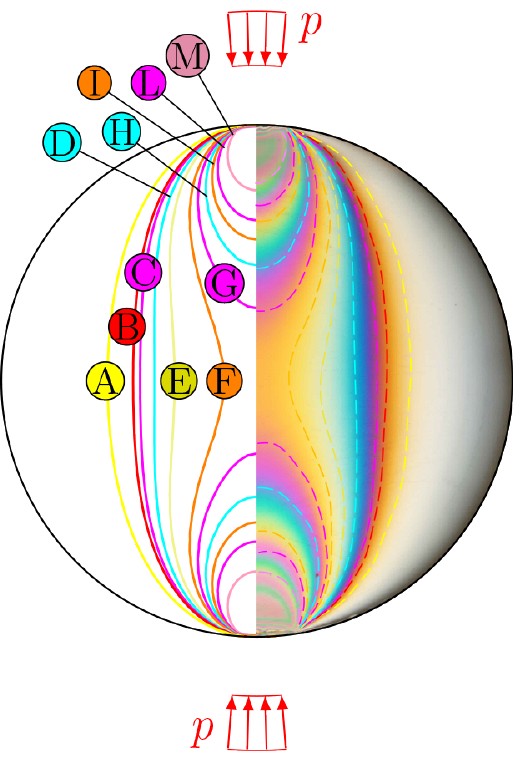}
	\caption{
	Left: A polymethyl methacrylate disk (subject to two opposite forces $F=7.1$kN) exhibits temporary birefringence displaying a fringe number $N=2$ near the centre, where the corresponding indigo-violet fringes are visible. 
	Right: the state of in-plane deviatoric stress (dashed line) compared with the theoretical results pertaining to the reported load distribution. Observation has been carried with a circular polariscope at a white light; quantitative results are reported in Table \ref{tabDisk}.
	}
	\label{PhDisk}
\end{figure}

Polimethyl Metacrilate, as other materials, exhibits the phenomenon known as \textit{temporary birefringence} \cite{frocht1965photoelasticity}.
The photoelastic parameters for our experiments have been determined by testing under 
diametral compression a disk  with radius 
$R = $~35~mm and thickness 
$b = $~5~mm. 
The usual identification of the  photoelastic parameters 
is  performed by using the elastic solution for a 
disk subject to opposite concentrated forces, for which 
the in-plane deviatoric stress $\left|\sigma_{I}-\sigma_{II}\right|$ at the centre of the disk is known \cite{dally1991experimental}. 
In this way, the photoelastic constant $f_{\sigma}$ of the material can be evaluated, once the fringe number $N$ at the centre of the disk has been determined. From a photoelastic experiment performed on the uncoated disk (reported in the lower part of Fig. \ref{sample_Geom}), the fringe number $N$ appearing at the centre of the disk is found to be $N=2$ from the left part of Fig. \ref{PhDisk}, referred to an applied force $F=7.1$ kN. Therefore, 
the photoelastic constant of the material is found to be approximately $f_{\sigma}=129.1$ N/mm.

\begin{table}[ht!]
		\begin{minipage}{\linewidth}
		\centering
		\begin{tabular}[h]{c@{\hspace{4\tabcolsep}}c@{\hspace{4\tabcolsep}}c@{\hspace{4\tabcolsep}}c@{\hspace{4\tabcolsep}}}\toprule[1.5pt]
			\bf Line & \bf Transmitted colour & \bf N$\lambda$ & $\bf\left|\sigma_{I}-\sigma_{II}\right|/\mu$ \\
			$[-]$       & $[-]$          &  $[\text{nm}]$ & $\left[-\right]$ \\\midrule
			A	& Brilliant yellow & 390 & 0.0148\\[5pt]
			B	& Red & 536 & 0.0203\\[5pt]
			C	& Indigo-violet & 575 & 0.0219\\[5pt]
			D	& Sky blue & 664 & 0.0252\\[5pt]
			E	& Brilliant yellow & 2$\cdot$390 & 0.0296\\[5pt]
			F	& Orange & 2$\cdot$505 & 0.0384\\[5pt]
			G	& Indigo-violet & 2$\cdot$575 & 0.0438\\[5pt]
			H	& Sky blue & 2$\cdot$664 & 0.0505\\[5pt]
			I	& Orange & 3$\cdot$505 & 0.0576\\[5pt]
			L	& Indigo-violet & 3$\cdot$575 & 0.0657\\[5pt]
			M	& purple & 4$\cdot$565 & 0.0859\\[5pt]
			\bottomrule
		\end{tabular}\par
		\captionof{table}{
		Determination of the in-plane deviatoric stress $\left|\sigma_{I}-\sigma_{II}\right|/\mu$ from the photoelastic experiment performed in the uncoated disk sample reported in the lower part of Fig. \ref{sample_Geom}.
		}
		\label{tabDisk}
		\end{minipage}
	\end{table}

Having evaluated $f_{\sigma}$, the in-plane deviatoric stress inside the disk can be evaluated for each integer fringe number $N$ (corresponding to indigo-violet transition  colour, visible at a white light). In order 
to determine the order of fractional fringes, the  in-plane deviatoric stress has to be  evaluated for a particular transmitted colour, once  its wave length $\lambda$ is known. In Table \ref{tabDisk} the dimensionless in-plane deviatoric stress for the coloured fringes are reported, as obtained  during the photoelastic experiment shown in Fig. \ref{PhDisk}
(dashed line in the right part), contrasted with the analytic results (obtained for the sketched traction distribution and drawn as continuous lines) obtained from  \cite{muskhelishvili2013some}.
	    \newpage
	\section{The ring, an annular beam \label{anelloApp}}
	The problem of a loaded circular beam is a classical exercise in structural mechanics and can completely be solved by means of the relations \eqref{azione}, once the expressions for the radial displacement $u_{r}$ in equation \eqref{urComp} is known in terms of the $A_{\pm{n}}$ coefficients. The stress distribution inside the circular beam can be determined once the internal forces have been determined from equations \eqref{beamAzione} in the limit $\mu\to{0}$ for the disk.
		\begin{figure}[htb!]
			\centering
            \includegraphics[scale=1]{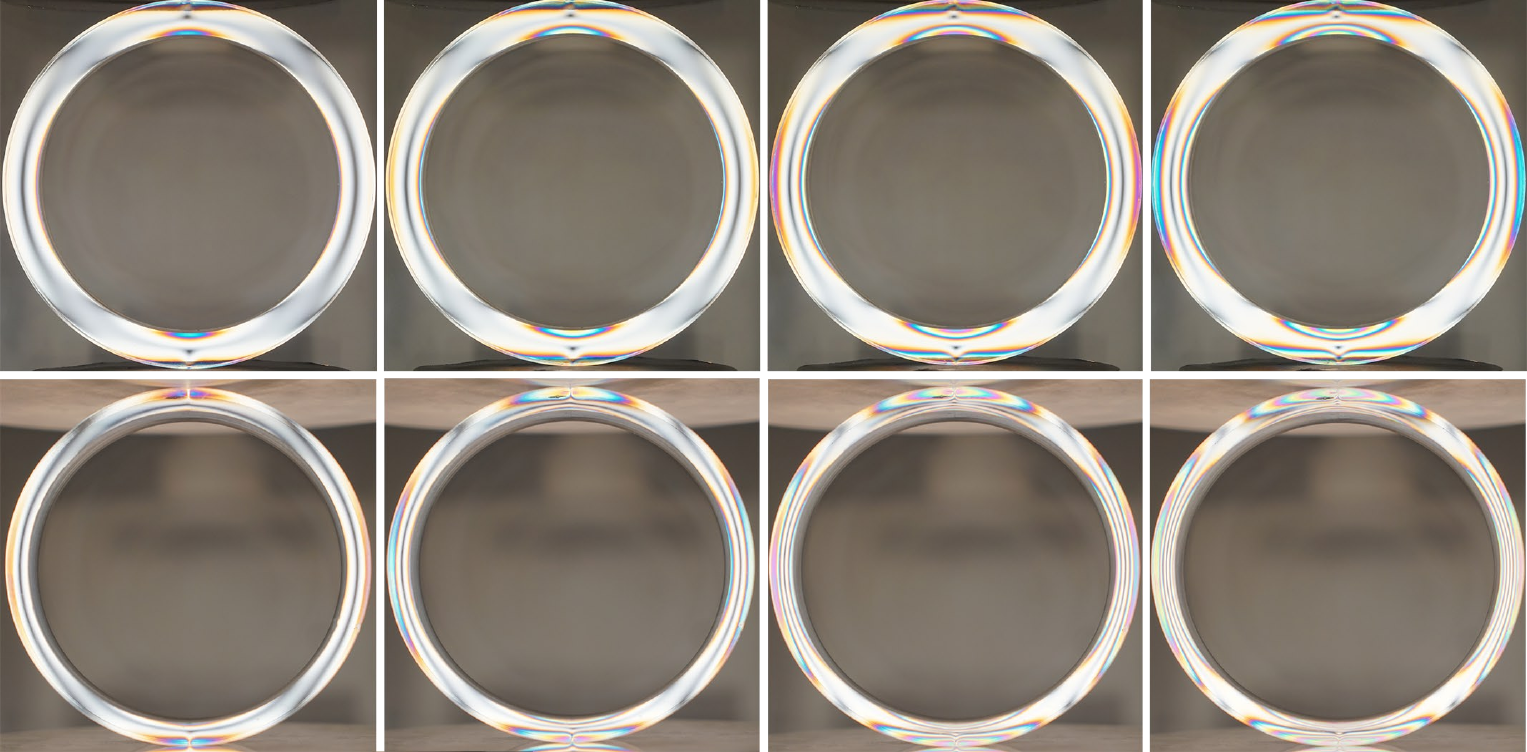}
			\caption{Photoelastic fringes generated during (vertical) diametrical compression of an annular beam, with bending stiffness EJ$_{1}$=1.4\,$\cdot$10$^{-2}$\,kNm$^{2}$ (upper part) and EJ$_{2}$=7\,$\cdot$10$^{-4}$\,kNm$^{2}$ (lower part). Four increasing values of compression are reported (from left to right): 0.30 kN (0.06 kN), 0.40 kN (0.15 kN), 0.50 kN (0.30 kN), and 0.60 kN (0.40 kN) in the upper part (in the lower part) for the samples shown in Fig. \ref{sample_Geom}; photos have been taken at white circularly polarized light.}
			\label{photoelasticity_anello}
		\end{figure}
	
\begin{figure}[hbt!]
		\centering
		\includegraphics[keepaspectratio,scale=1.15]{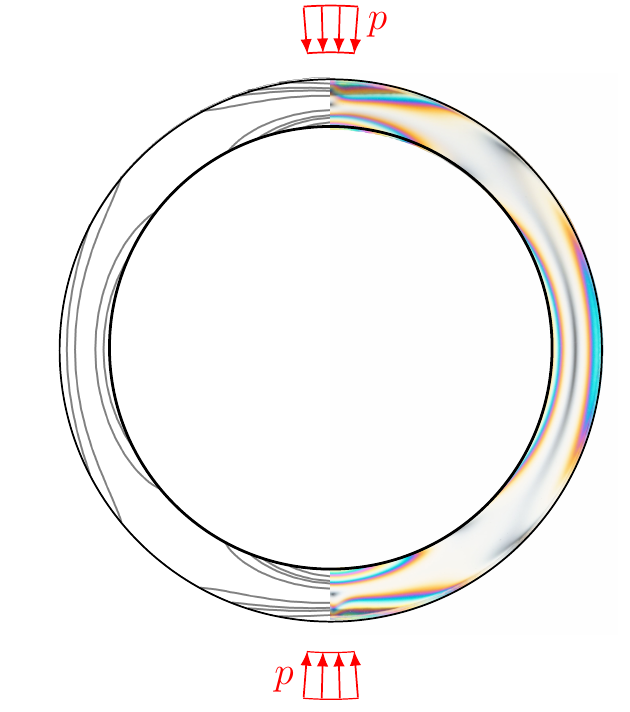}
		\hspace{0.5 pt}
		\includegraphics[keepaspectratio,scale=1.15]{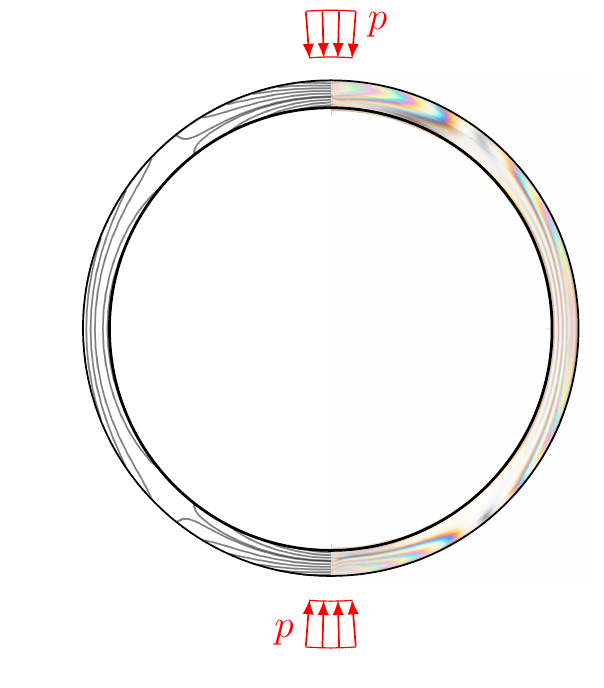}
		\caption{Distribution of the normalized in-plane deviatoric stress  $(\sigma_{I}-\sigma_{II})/EJ_{j}$ (j=1,2) compared with the photoelastic fringes during a diametral compression test of two photoelastic circular beams with  bending stiffness  EJ$_{1}$=1.4\,$\cdot$10$^{-2}$\,kNm$^{2}$ (left) and EJ$_{2}$=7\,$\cdot$10$^{-4}$\,kNm$^{2}$ (right). The applied external load distribution $p$ is modelled with a Fourier series expansion  truncated at $N=100$ and enhanced with the Lanczos smoothing method. The load distribution results in a compression of 0.60 kN (left) and 0.40 kN (right).
		}
		\label{anello_ph}
	\end{figure}

	Photoelastic experiments on the annular beams shown in Fig. \ref{sample_Geom} are reported in Fig. \ref{photoelasticity_anello}. 
	The photos refer to four different compressive values of load, namely, 0.30 kN (0.06 kN), 0.40 kN (0.15 kN), 0.50 kN (0.30 kN), and 0.60 kN (0.40 kN) in the upper part (in the lower part).
    Maps of the normalized in-plane deviatoric stress $\left|\sigma_{I}-\sigma_{II}\right|/EJ$ are compared in Fig. \ref{anello_ph} with the photoelastic experiments, loaded at 0.60 kN and 0.40 kN. 

	The results demonstrate the high accuracy of the beam theory to produce the stress distribution in a circular beam. This is a known result and can be found in \cite{frocht1965photoelasticity}.
%
\end{appendices}
	{\newpage
	\renewcommand{\bibname}{References}
	\bibliographystyle{ieeetr} 

\end{document}